\documentclass[twocolumn,
prb,
showpacs,
superscriptaddress,
floatfix
]{revtex4-1} 

\usepackage[dvipdfmx]{graphicx}
\usepackage{epsf}
\usepackage{dcolumn}
\usepackage{bm}
\usepackage{ulem}
\usepackage{amsmath}
\usepackage{amssymb}
\usepackage{color}

\newcommand{\tr}  {{\rm tr}}
\usepackage[pdftex,
colorlinks=true,
citecolor=blue,
setpagesize=false]{hyperref}

\begin{document}

\title{
Topological property of a $t_{2g}^5$ system with a honeycomb lattice structure
}

\author {Beom Hyun Kim}
\affiliation{Korea Institute for Advanced Study, Seoul 02455, South Korea}
\affiliation{
  Computational Condensed Matter Physics Laboratory, 
  RIKEN Cluster for Pioneering Research (CPR), 
  Saitama 351-0198, Japan}
\author {Kazuhiro Seki}
\affiliation{
International School for Advanced Studies (SISSA), 
Via Bonomea 265, 34136, Trieste, Italy
}
\affiliation{
  Computational Condensed Matter Physics Laboratory, 
  RIKEN Cluster for Pioneering Research (CPR), 
  Saitama 351-0198, Japan}
\affiliation{
  Computational Materials Science Research Team, 
  RIKEN Center for Computational Science (R-CCS),  
  Kobe, Hyogo 650-0047, Japan}
\author {Tomonori Shirakawa}
\affiliation{
International School for Advanced Studies (SISSA), 
Via Bonomea 265, 34136, Trieste, Italy
}
\affiliation{
  Computational Condensed Matter Physics Laboratory, 
  RIKEN Cluster for Pioneering Research (CPR), 
  Saitama 351-0198, Japan}
\affiliation{
  Computational Materials Science Research Team, 
  RIKEN Center for Computational Science (R-CCS),  
  Kobe, Hyogo 650-0047, Japan}
\affiliation{Computational Quantum Matter Research Team, RIKEN, 
  Center for Emergent Matter Science (CEMS), Wako, Saitama 351-0198, Japan}
\author {Seiji Yunoki}
\affiliation{
  Computational Condensed Matter Physics Laboratory, 
  RIKEN Cluster for Pioneering Research (CPR), 
  Saitama 351-0198, Japan}
\affiliation{
  Computational Materials Science Research Team, 
  RIKEN Center for Computational Science (R-CCS),  
  Kobe, Hyogo 650-0047, Japan}
\affiliation{Computational Quantum Matter Research Team, RIKEN, 
  Center for Emergent Matter Science (CEMS), Wako, Saitama 351-0198, Japan}

\date{\today}

\begin{abstract}
A $t_{2g}^5$ system with a honeycomb lattice structure such as Na$_2$IrO$_3$
was firstly proposed as a topological insulator 
even though Na$_2$IrO$_3$ and its isostructural materials in nature have been
turned out to be a Mott insulator with magnetic order. 
Here we theoretically revisit the topological property based on a minimal 
tight-binding Hamiltonian for three $t_{2g}$ bands incorporating a strong
spin orbit coupling and two types of the first nearest neighbor (NN) 
hopping channel between transition metal ions, i.e., 
the hopping ($t_1$) mediated by edge-shared ligands and the direct hopping ($t_1'$) 
between $t_{2g}$ orbitals via $dd\sigma$ bonding.
We demonstrate that the topological phase transition takes place
by varying only these hopping parameters with the relative strength parametrized by $\theta$,  
i.e., $t_1=t\cos\theta$ and $t_1'=t\sin\theta$. 
We also explore the effect of the second and third NN hopping channels, 
and the trigonal distortion on the topological phase for the whole range of $\theta$. 
Furthermore, we examine the electronic and topological phases in the
presence of on-site Coulomb repulsion $U$.
Employing the cluster perturbation theory, we show that, with increasing $U$, 
a trivial or topological band insulator in the absence of $U$
can be transferred into
a Mott insulator with nontrivial or trivial band topology. 
We also show that the main effect of the Hund's coupling 
can be understood simply as the renormalization of $U$.
We briefly discuss the relevance of our results to the existing materials. 
\end{abstract}


\maketitle

\section{Introduction}

Topology of electronic states is one of the most fascinating 
research subjects in the current condensed matter physics.
This is a new physical aspect to distinguish quantum phases
beyond the traditional Landau's approach based on the spontaneous symmetry breaking. 
The quantum spin Hall (QSH) phase, 
which can arise in the presence of the time-reversal symmetry (TRS),
is the most extensively studied example of intriguing topological phases~\cite{Hasan2010}.
The QSH insulator, termed as a topological insulator (TI), is 
characterized by the $Z_2$ topological invariant 
determined by the time-reversal polarization~\cite{Fu2006,Fu2007a,Fu2007b}.
In contrast with a conventional insulator, gapless edge or surface states 
protected by the TRS emerge along with a peculiar magnetoelectric effect~\cite{Ando2013}.
After the theoretical proposal of the QSH phase in graphene~\cite{Kane2005a,Kane2005b} 
and HgTe quantum well~\cite{Bernevig2006},
many theoretical and experimental researches have verified that 
not only a TI but also other types of topological phases such as a topological crystalline 
insulator and a Weyl semimetal are indeed stabilized in existing materials~\cite{Bansil2016}.

TIs and many candidate TIs are $5p$- or $6p$-based 
with a strong spin-orbit coupling (SOC) such as Bi$_2$Se$_3$~\cite{Xia2009,Zhang2009}, 
and only a few candidates have been proposed in 
$4d$ or $5d$ transition metal (TM) compounds~\cite{Shitade2009,Pesin2010,
Yang2010,CHKim2012,HSKim2013,Qian2014,Weng2015,Sun2015,Zhou2015,Ochi2016,Khazaei2016,Si2016}.
Na$_2$IrO$_3$ is the first candidate of TM-based TIs.
This system is in the low-spin state of Ir$^{4+}$ ion, stabilized due to
the gigantic cubic crystal field of approximately 3 eV, 
with five electrons per TM occupying in Ir $t_{2g}$-based bands, which
are split into four-fold degenerate $j_{\rm eff}=3/2$ bands and two-fold degenerate 
$j_{\rm eff}=1/2$ bands in the presence of a strong SOC. 
Here, $j_{\rm eff}$ is referred to as the effective total angular momentum. 
Because of their large splitting, 
it is expected that the four-fold degenerate $j_{\rm eff}=3/2$ bands are fully occupied, and 
only the doubly degenerate $j_{\rm eff}=1/2$ bands cross the Fermi energy and are half filled. 
Because Ir atoms in Na$_2$IrO$_3$ form a layered honeycomb lattice and 
the energy band dispersion along the inter-layer direction is much smaller 
than that in the intra-layer plane, 
the low-energy electronic structure of Na$_2$IrO$_3$ can be mapped into 
an effective tight-binding model for the $j_{\rm eff}=1/2$ bands, which is reminiscent of
the Kane-Mele model of graphene~\cite{Kane2005a,Kane2005b}.  
If the parameters in the effective Kane-Mele model
is in the right range, the QSH phase necessarily emerges in Na$_2$IrO$_3$. 
This point has been firstly pointed out by Shitade {\it et al.}~\cite{Shitade2009}.
The consecutive studies have supported a weak TI in Na$_2$IrO$_3$ and 
a strong TI in isostructural Li$_2$IrO$_3$ 
if the trigonal distortion and further neighbor hoppings 
are tailored suitably~\cite{CHKim2012,HSKim2013}.

In spite of the theoretical prediction, the QSH phase in Na$_2$IrO$_3$ and 
its isostructural materials, Li$_2$IrO$_3$, Li$_2$RhO$_3$, and $\alpha$-RuCl$_3$, 
has not been experimentally reported yet.
In fact, these materials prefer to exhibit topologically trivial insulating phases 
with long-range magnetic order~\cite{Singh2010,Singh2012,Luo2013,Sears2015}.
These phases are rather understood in terms of the Mott physics of relativistic $d$ orbitals 
with a strong SOC~\cite{Comin2012,MJKim2016,BHKim2016}.
Although the spatial distribution of $4d$ or $5d$ orbitals is somewhat 
extended as compare with that of $3d$ orbitals, 
the electron correlation could be hardly screened out 
and still play a role in determining their electronic characteristics.
There have also been extensive studies along this line on these systems, focusing on 
their exotic magnetic phases induced by the mutual interplay among the kinetic energy, 
SOC, and Coulomb interaction, which include, for example, Kitaev spin liquid 
phase~\cite{Chaloupka2010,Jiang2011,Reuther2011}.

Nevertheless, the possibility of the QSH phase in these systems is still interesting. 
State-of-the-art structural control with pressure, 
chemical substitution, or substrate engineering can potentially manipulate experimentally 
their electronic kinetics and correlations. 
Moreover, a recent photoemission spectroscopy experiment has observed metallic 
surface states near the $\Gamma$ point in Na$_2$IrO$_3$~\cite{Alidoust2016,
Moreschini2017}. 
Despite that its texture is not direct evidence on the QSH phase, 
in which a gapless mode is expected to appear at the $M$ points, 
it still infers the possibility 
that the electronic character near surfaces could be quite different 
from the bulk Mott insulating phase~\cite{Alidoust2016,Moreschini2017}.

The topological phase transition in the presence of electron 
correlations has also attracted much attention.
A lot of theoretical approaches have been employed to determine
electronic and topological phases 
of various interacting topological insulators~\cite{Rachel2018}.
When the correlations are weak, both electronic and topological phases 
are still robust in the topological band insulator (TBI) 
even though the insulating gap can be slightly modified.
In the limit of strong electron correlations, an electronic phase is surely changed
from a band insulator (BI) to a Mott insulator (MI) with often magnetic order. 
In a moderate correlation regime, however, exotic electronic and
topological phases are expected. 
The mean-field approximation based on the slave-rotor approach has shown 
the possibility of a topological Mott insulator (TMI) in pyrochlore iridates~\cite{Pesin2010}.
The first-principles electronic structure calculation~\cite{Wan2011} and 
the cellular dynamical mean-field theory (CDMFT)~\cite{AGo2012} 
have revealed that exotic topological phases such as an axion insulator and 
a Weyl semimetal emerge between a TI and an antiferromagnetic (AFM) MI 
when the electron correlations are increased. 
The topological phase transition of the Kane-Mele-Hubbard model 
for interacting graphene has also been investigated by 
various numerical methods~\cite{Hohenadler2011,Yu2011,Hohenadler2012,Wu2012,Hohenadler2014,Grandi2015}.
The effect of correlations in the effective $j_{\rm eff}=1/2$ model 
proposed by Shitade {\it et al}. has also been studied with the slave-spin approaches~\cite{Ruegg2012}.

All these studies have found that the topological phase transition occurs 
from a TBI to an AFM MI with increasing the Coulomb repulsion. 
However, the contradicting results are obtained among the different studies 
on the nature of the intermediate phase. 
The CDMFT calculations~\cite{Wu2012} have found a spin liquid phase 
near the phase boundary in a very weak SOC region, whereas 
the quantum Monte Carlo (QMC) method~\cite{Hohenadler2014} and the cluster perturbation theory (CPT) method~\cite{Grandi2015}
have not predicted the presence of the spin liquid phase.
The single-particle excitation gap is perfectly closed at the critical point in the calculations using the CPT and 
variational cluster approximation (VCA) methods~\cite{Yu2011,Grandi2015}. 
In contrast, the QMC calculations show that the gap becomes smallest 
but remains finite~\cite{Hohenadler2011,Hohenadler2012}. 
Recent studies on the Haldane-Hubbard model using 
the VCA have found nonmagnetic and magnetic TBI phases 
in the presence of electron correlations~\cite{Gu2015,Wu2016}.

Here, in this paper, we revisit the topological property of 
Na$_2$IrO$_3$ and its isostructural compounds theoretically by 
considering a $t_{2g}^5$ system in the single-layer honeycomb lattice.
In contrast with the previous studies, which mainly elucidate the role of 
longer-range hoppings, i.e., the second nearest-neighbor (2nd NN) and 
third nearest-neighbor (3rd NN) hopping channels, in the topological phase in 
the analogy of the Kane-Mele model, we focus on the 
two dominant processes in the first nearest-neighbor (1st NN) hopping channel between TMs: 
the direct $d$-$d$ hopping via the $dd\sigma$ bonding and the indirect
hopping mediated by edge-shared ligands via the $pd\pi$ bonding, 
and examine the topological phase transition. 
We demonstrate that the topological phase is transferred from a trivial
BI to a TBI and vice versa with varying the relative strength between the two different hoppings 
in the 1st NN channel.
In addition, we explore the topological phase transition against the Coulomb repulsion. 
Employing the CPT, we calculate the electronic and
topological phase diagram in the presence of the Coulomb repulsion. 
We find that
a Mott insulator with nontrivial band topology similar to the QSH state
appears over a broaden parameter region of the phase diagram.

The rest of this paper is organized as follows. 
Section~\ref{sec:method} introduces a model Hamiltonian of 
the $t_{2g}$ system and explains briefly numerical methods to calculate 
the topological invariant for both noninteracting and interacting cases. 
The topological phase diagram in the noninteracting limit 
with respect to the SOC and the 1st NN hopping parameters is 
examined in Sec.~\ref{sec:non-int}. 
The edge states in a zigzag stripy geometry are also 
analysed. 
The roles of the trigonal distortion as well as the 2nd and 3rd NN hopping channels 
in the topological phase is also studied in Sec.~\ref{sec:non-int}. 
The effect of electron correlations on the topological phase diagram 
is investigated in Sec.~\ref{sec:inter}. 
Finally, Sec.~\ref{sec:discussion} discusses the relevance of our results to the existing $t_{2g}^5$ compounds, 
before concluding the paper in Sec.~\ref{sec:conclusion}. 
Appendix~\ref{app:cpt} provides the details of the CPT used here, followed by the results of the 
single-particle excitation spectrum  in Appendix~\ref{app:sf} and the topological Hamiltonian in 
Appendix~\ref{app:th}

\section{Model and Method}\label{sec:method}

\subsection{Noninteracting Hamiltonian}

\begin{figure}[b]
\centering
\includegraphics[width=.95\columnwidth]{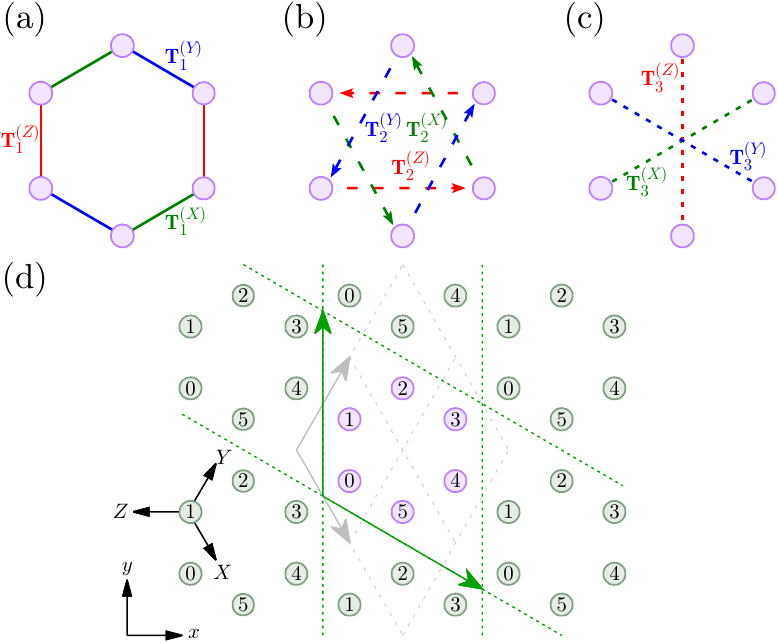}
\caption{
 (a--c) Schematic diagrams describing three types of hopping channels, i.e., 
 (a) $\mathbf{T}^{(\gamma)}_1$ for the 1st NN hopping, (b) $\mathbf{T}^{(\gamma)}_2$ for the 2nd NN hopping, 
 and (c) $\mathbf{T}^{(\gamma)}_3$ for the 3rd NN hopping, where $\gamma\,(=X,Y$, and $Z$) distinguishes three 
 different bonds indicated by different colors for each type of hopping channels. 
 Note that $\mathbf{T}^{(\gamma)}_2$ depends also 
 on the hopping direction and the $\gamma$ direction is defined by arrows in (b). The opposite hopping direction to the 
 $\gamma$ direction is denoted as the $\bar\gamma$ direction in the text.   
 (d) A schematic honeycomb lattice structure divided into supercell clusters, where each supercell cluster is 
 composed of six sites enumerated from 0 to 5. 
 Gray and green arrows refer to basis vectors of the original honeycomb lattice and
 the honeycomb lattice composed of the supercell clusters, respectively.
 All lattice sites are laid on the $xy$-plane and the $z$ direction is perpendicular to the plane. 
 $X$, $Y$, and $Z$ are mutually orthogonal local coordinates to define the $t_{2g}$ orbitals, i.e.,  
 $d_{XY}$, $d_{YZ}$, and $d_{ZX}$ orbitals.
 Unit vectors of the local coordinates are given as 
 $\hat{X}=\sqrt{\frac{1}{6}}\hat{x}-\sqrt{\frac{1}{2}}\hat{y}
 +\sqrt{\frac{1}{3}}\hat{z}$, 
 $\hat{Y}=\sqrt{\frac{1}{6}}\hat{x}+\sqrt{\frac{1}{2}}\hat{y}
 +\sqrt{\frac{1}{3}}\hat{z}$, and
 $\hat{Z}=-\sqrt{\frac{2}{3}}\hat{x}+\sqrt{\frac{1}{3}}\hat{z}$, where 
 $\hat x$, $\hat y$, and $\hat z$ are unit vectors of the global coordinates 
 indicated in the figure. 
}
\label{Fig_HC}
\end{figure}

To investigate the electronic and topological phases of a $t_{2g}^5$ system
with the honeycomb lattice structure such as Na$_2$IrO$_3$ and its isostructural systems,
we consider three hopping channels between 1st NN, 2nd NN, 
and 3rd NN sites, as schematically shown in Figs.~\ref{Fig_HC}(a)--\ref{Fig_HC}(c).
Let $\mathbf{T}^{(\gamma)}_1$, $\mathbf{T}^{(\gamma)}_2$, and 
$\mathbf{T}^{(\gamma)}_3$ be the $3\times3$ hopping matrices of $\gamma$-type 
($\gamma=X$, $Y$, and $Z$) for the 1st, 2nd, and 3rd NN hoppings, respectively.
Because there is no inversion symmetry (IS) about the bond center of sites connected via the 2nd NN hopping 
[see Fig.~\ref{Fig_HC}(b)], 
$\mathbf{T}^{(\gamma)}_2$ along the $\gamma$ direction, indicated by arrows in Fig.~\ref{Fig_HC}(b), 
is not the same as that along the opposite direction denoted as $\bar{\gamma}$. 
The hopping matrix along the opposite hopping direction, $\mathbf{T}^{(\bar{\gamma})}_2$, is given by 
the transpose of $\mathbf{T}^{(\gamma)}_2$. 
In contrast, the other two hopping matrices $\mathbf{T}^{(\gamma)}_1$ and $\mathbf{T}^{(\gamma)}_3$ 
are independent of the hopping directions because there is the IS at the center of the corresponding bond, 
and hence $\mathbf{T}^{(\bar\gamma)}_{1(3)} = \mathbf{T}^{(\gamma)}_{1(3)}$.

We consider the following tight-binding Hamiltonian $H_t$ on the honeycomb lattice: 
\begin{align}
H_t &= 
  \sum_{ i, \gamma, \alpha, \beta, \sigma} T^{(\gamma)}_{1,\alpha \beta} 
 c_{i_{1\gamma} \alpha \sigma}^{\dagger} c_{i \beta \sigma}
 + \sum_{ i, \gamma, \alpha, \beta, \sigma} T^{(\gamma)}_{2,\alpha \beta} 
 c_{i_{2\gamma} \alpha \sigma}^{\dagger} c_{i \beta \sigma} \nonumber \\
 &+ \sum_{ i, \gamma, \alpha, \beta, \sigma} T^{(\bar{\gamma})}_{2,\alpha \beta} 
 c_{i_{2\bar{\gamma}} \alpha \sigma}^{\dagger} c_{i \beta \sigma} 
 + \sum_{ i, \gamma, \alpha, \beta, \sigma} T^{(\gamma)}_{3,\alpha \beta} 
 c_{i_{3\gamma} \alpha \sigma}^{\dagger} c_{i \beta \sigma} \nonumber \\
 &+\lambda \sum_{i,\alpha,\beta,\sigma,\sigma^{\prime}}
   (\mathbf{l}\cdot\mathbf{s})_{\alpha\sigma,\beta\sigma^{\prime}} 
   c_{i\alpha\sigma}^{\dagger}c_{i\beta\sigma^{\prime}} 
 - \mu_{t} \sum_{i, \alpha, \sigma} 
   c_{i \alpha \sigma}^{\dagger} c_{i \alpha \sigma}  \nonumber \\
  &+ \frac{\Delta_{\rm tr}}{3}  \sum_{i, \sigma} \Big( 
    c_{i\tilde{x}\sigma}^{\dagger} c_{i\tilde{x}\sigma}
  + c_{i\tilde{y}\sigma}^{\dagger} c_{i\tilde{y}\sigma} 
  -2 c_{i\tilde{z}\sigma}^{\dagger} c_{i\tilde{z}\sigma}
  \Big),
\end{align}
where $c_{i\alpha\sigma}$ is the annihilation operator of electron with
orbital $\alpha\,(=XY$, $YZ$, and $ZX$) and spin $\sigma\,(=\pm\frac{1}{2})$ 
at lattice site $i$, and $XY$, $YZ$, and $ZX$ are three $t_{2g}$ orbitals, i.e., 
$d_{XY}$, $d_{YZ}$, and $d_{ZX}$ orbitals, 
represented in the local coordinates indicated in Fig.~\ref{Fig_HC}(d) (also see
Ref.~\onlinecite{BHKim2016}). The first four terms describe the electron hopping, where 
$i_{1\gamma}$, $i_{2\gamma}$, and $i_{3\gamma}$ are site indices, denoting sites connected 
from site $i$ via the $\gamma$-type 1st, 2nd, and 3rd NN hopping channels, 
respectively, with $\gamma=X$, $Y$, and $Z$, as shown in Figs.~\ref{Fig_HC}(a)--\ref{Fig_HC}(c). 
Note that each site has three neighboring sites that are connected via the 1st and 3rd NN hopping channels, 
while there are six neighboring sites that are connected from a given site via the 2nd NN hopping channel.

The fifth term in $H_t$ is the SOC Hamiltonian and the matrix elements of $\mathbf{l}\cdot\mathbf{s}$ 
are given as $\left( \mathbf{l}\cdot\mathbf{s} \right)_{\alpha\sigma,\beta\sigma'}
=\langle \alpha|\mathbf{l}|\beta\rangle \cdot \langle\sigma|\mathbf{s}|\sigma'\rangle$, 
where $\mathbf{l}$ and $\mathbf{s}$ are orbital and spin angular momentum
operators, respectively. Among these matrix elements, the nonzero matrix elements are 
$\langle XY,\pm\frac{1}{2}|\mathbf{l}\cdot\mathbf{s}|YZ,\mp\frac{1}{2}\rangle=\pm\frac{1}{2}$, 
$\langle YZ,\pm\frac{1}{2}|\mathbf{l}\cdot\mathbf{s}|ZX,\pm\frac{1}{2}\rangle=\pm \frac{i}{2}$, 
$\langle ZX,\pm\frac{1}{2}|\mathbf{l}\cdot\mathbf{s}|XY,\mp\frac{1}{2}\rangle=\frac{i}{2}$,
and the complex conjugate of these elements. 
The SOC causes the six-fold degenerate $t_{2g}$ orbitals, including the spin degree of freedom, 
to split into four-fold degenerate 
$j_{\rm eff}=3/2$ and doubly degenerate $j_{\rm eff}=1/2$ relativistic orbitals. 
$\mu_t$ in the sixth term in $H_t$ is the chemical potential and is determined 
for the number of electrons per site to be $5$.

The last term in $H_t$ describes the energy level splitting due to the trigonal distortion.
In the presence of the trigonal distortion, the three-fold degenerate $t_{2g}$ orbitals, not including 
the spin degree of freedom, split into  
doubly degenerate $e_g'$ ($\tilde{x}$ and $\tilde{y}$) orbitals and
nondegenerate $a_{1g}$ ($\tilde{z}$) orbital with
the level splitting energy $\Delta_{\rm tr}=E_{\tilde{x}(\tilde{y})}-E_{\tilde{z}}$.
Here, $\tilde{x}$, $\tilde{y}$ and $\tilde{z}$ orbitals are given as 
$|\tilde{x}\rangle=\frac{1}{\sqrt{6}}\left(|ZX\rangle-2|XY\rangle+|YZ\rangle\right)$,
$|\tilde{y}\rangle=\frac{1}{\sqrt{2}}\left(|ZX\rangle-|YZ\rangle\right)$, and
$|\tilde{z}\rangle=\frac{1}{\sqrt{3}} \left(|ZX\rangle+|XY\rangle+|YZ\rangle\right)$.
The SOC term has nonzero matrix elements
in these $\tilde{x}$, $\tilde{y}$, and $\tilde{z}$ orbitals only for 
$\langle \tilde{x},\pm\frac{1}{2}|\mathbf{l}\cdot\mathbf{s}
|\tilde{y},\pm \frac{1}{2} \rangle=\pm\frac{\textrm{i}}{2}$,
$\langle \tilde{y},\pm\frac{1}{2}|\mathbf{l}\cdot\mathbf{s}
|\tilde{z},\mp \frac{1}{2} \rangle=\frac{\textrm{i}}{2}$, 
$\langle \tilde{z},\pm\frac{1}{2}|\mathbf{l}\cdot\mathbf{s}
|\tilde{x},\mp \frac{1}{2} \rangle=\pm\frac{1}{2}$, 
and the complex conjugate of these elements.

For simplicity, we only consider one or two hopping processes in each hopping channel, which contribute
dominantly for the hopping channel, as previously estimated 
in Refs.~\onlinecite{CHKim2012,Foyevtsova2013,Yamaji2014,Winter2016}. 
The hoppings considered here in this study is summarized in Table~\ref{hopM}. 
For the 1st NN hopping channel, $t'_1$ and $t_1$ refer to the hopping amplitudes of 
the direct hopping via the $dd\sigma$ bonding of $d$ orbitals 
and the indirect hopping mediated via the $pd\pi$ bonding between a TM and its neighboring ligands, respectively. 
These hopping amplitudes $t_1$ and $t'_1$ are parametrized as 
$t_1=t\cos\theta$ and $t_1'=t\sin\theta$ by 
introducing two parameters $t\,(>0)$ and $\theta$. 
As already noted above, the broken IS of the 2nd NN hopping channel gives rise to different hopping 
amplitudes $t_2$ from orbital $\alpha$ to orbital $\beta$ and $t_2'$ from orbital $\beta$ to orbital $\alpha$. 
Because the previous studies have supported that both 2nd and 3rd NN hoppings 
are negative~\cite{CHKim2012,Foyevtsova2013,Yamaji2014,Winter2016}, 
here we only consider negative $t_2$, $t_2'$, and $t_3$.

\begin{table}[b]
\centering
\caption{
Non-zero hopping matrix elements for the 1st, 2nd, and 3rd NN hopping channels, 
$\mathbf{T}_1^{(\gamma)}$, $\mathbf{T}_{2}^{(\gamma)}$, and 
$\mathbf{T}_{3}^{(\gamma)}$, of $\gamma\,(\,=X,Y$, and $Z)$ type.
}
\label{hopM}
\begin{ruledtabular}
\begin{tabular} { c c c c }
$\gamma$ &
$\mathbf{T}_1^{(\gamma)}$ &
$\mathbf{T}_{2}^{(\gamma)}$ &
$\mathbf{T}_{3}^{(\gamma)}$ \\
\hline
$X$ & 
\begin{tabular}{c} $t_1':YZ \rightarrow YZ$ \\ 
                   $t_1: ZX \rightarrow XY$ \\ 
                   $t_1: XY \rightarrow ZX$ \end{tabular} &
\begin{tabular}{c} $t_2':ZX \rightarrow XY$ \\ 
                   $t_2: XY \rightarrow ZX$ \end{tabular} &
$t_3 : YZ \rightarrow YZ$  \\
\hline
$Y$ & 
\begin{tabular}{c} $t_1':ZX \rightarrow ZX$ \\ 
                   $t_1: XY \rightarrow YZ$ \\ 
                   $t_1: YZ \rightarrow XY$ \end{tabular} &
\begin{tabular}{c} $t_2':XY \rightarrow YZ$ \\ 
                   $t_2: YZ \rightarrow XY$ \end{tabular} &
$t_3 : ZX \rightarrow ZX$  \\
\hline
$Z$ & 
\begin{tabular}{c} $t_1':XY \rightarrow XY$ \\ 
                   $t_1: YZ \rightarrow ZX$ \\ 
                   $t_1: ZX \rightarrow YZ$ \end{tabular} &
\begin{tabular}{c} $t_2':YZ \rightarrow ZX$ \\ 
                   $t_2: ZX \rightarrow YZ$ \end{tabular} &
$t_3 : XY \rightarrow XY$  
\end{tabular}
\end{ruledtabular}
\end{table}

\subsection{$Z_2$ topological invariant}

Because $H_t$ possesses both IS and TRS simultaneously,
the topological characteristic induced by the TRS can be 
investigated without directly calculating the Berry curvature 
over the whole momentum space. 
Owing to the theory by Fu and Kane~\cite{Fu2007a},
the $Z_2$ topological invariant $\nu$ of the QSH phase can be
evaluated simply from parity eigenvalues of occupied energy bands
at every time-reversal invariant momentum (TRIM) point.
Provided that $(2m\!-\!1)$- and $2m$-th energy bands are $m$-th Kramers pair
(ascending order in energy eigenvalues)  
with the same energy and parity eigenvalues at a TRIM point, 
the topological quantity $(-1)^\nu$ is given as
\begin{equation}
(-1)^{\nu}=
\prod_{i=1}^{4} \delta_{\Lambda_i} =
\prod_{i=1}^{4} \prod_{m=1}^{n_v} \xi_{m} (\Lambda_i),
\label{Eq_nu}
\end{equation}
where 
$\xi_{m}(\Lambda_i)$ ($=\! \pm 1$) is the parity eigenvalue of the $m$-th Kramers pair
at specific TRIM $\Lambda_i$.
$n_v$ is total number of Kramers pairs below the Fermi energy 
and $n_v=5$ for our $t_{2g}^5$ system in the honeycomb lattice.
The topological quantity $\delta_{\Lambda_i}$ is defined as the product of
all parity eigenvalues of Kramers pairs below the Fermi energy at $\Lambda_i$.
For the honeycomb lattice, there are four TRIM points at the $\Gamma$ point and three $M$ points 
($M_1$, $M_2$, and $M_3$) [see red dots in Fig.~\ref{Fig_edge}(b)].

\subsection{Correlation effect}

To investigate the correlation effect, 
we add to $H_t$ the Kanamori-type interaction term described by the following Hamiltonian: 
\begin{align}
H_U &= \frac{1}{2}\sum_{i,\sigma,\sigma^{\prime},\alpha,\beta}
  U_{\alpha\beta} c_{i\alpha\sigma}^{\dagger}
  c_{i\beta\sigma^{\prime}}^{\dagger}
  c_{i\beta\sigma^{\prime}}c_{i\alpha\sigma}  \nonumber \\
  &+ \frac{1}{2}\sum_{i,\sigma, \sigma^{\prime},\alpha\ne\beta}
  J_{\alpha\beta}c_{i\alpha\sigma}^{\dagger}
  c_{i\beta\sigma^{\prime}}^{\dagger}
  c_{i\alpha\sigma^{\prime}}c_{i\beta\sigma}  \nonumber \\
  &+\frac{1}{2}\sum_{i,\sigma, \alpha\ne\beta}
  J_{\alpha\beta}^{\prime} c_{i\alpha\sigma}^{\dagger}
  c_{i\alpha\bar{\sigma}}^{\dagger}
  c_{i\beta\bar{\sigma}}c_{\beta\sigma},
\label{Eq:Corr}
\end{align}
where $U_{\alpha\alpha}=U$ and $U_{\alpha\beta}=U-2J_{\rm H}$ with $\alpha\ne\beta$ are 
the intra-orbital and inter-orbital on-site Coulomb interactions, respectively, 
$J_{\alpha\beta}=J_{\alpha\beta}^{\prime}=J_{\rm H}$ represents the Hund's coupling, and 
$\bar{\sigma}$ stands for the opposite spin of $\sigma$.
We employ the CPT of a six-site cluster depicted in Fig.~\ref{Fig_HC}(d)
to examine the electronic band structure of 
the interacting system by calculating the single-particle excitation spectrum~\cite{Senechal2008}.
According to a recent study~\cite{Grandi2015}, 
the symmetry of cluster is crucial to
determining the topological property in a honeycomb lattice 
because the discrepancy of the symmetry between the cluster and the original lattice 
leads to wrong symmetry of the self-energy in the single-particle Green's function, 
which can give rise to artificial electronic and topological phases.
Our selection of the cluster is the minimum cluster to keep the point group symmetry 
of the original honeycomb lattice.
Details on the CPT used here are described in appendix~\ref{app:cpt}.

To identify the topological property of an interacting system,
we adopt the framework proposed by Wang and Zhang~\cite{Wang2012a,Wang2012b}.
In this framework, the topological property of an interacting system
is evaluated from the corresponding noninteracting system described by the so-called 
``topological Hamiltonian'' $\mathbf{H}_{\rm T}(\mathbf{k})=-\mathbf{G}^{-1}(0,\mathbf{k})$, 
where $\mathbf{G}(\zeta,\mathbf{k})$ is the single-particle 
Green's function of the interacting system 
at frequency $\zeta$ and momentum $\mathbf{k}$.
This is justified because there always exists the smooth transformation from 
the single-particle Green's function of the noninteracting system described by  
the topological Hamiltonian, 
$\left[\zeta-\mathbf{H}_{\rm T}(\mathbf{k}) \right]^{-1}$, 
to that of the interacting system~\cite{Wang2012a}. 
We can calculate the topological invariant $(-1)^\nu$ of the interacting system
by using Eq.~(\ref{Eq_nu}), in which the parity eigenvalues are evaluated for the eigenstates of 
$\mathbf{H}_{\rm T}(\mathbf{k})$ with the negative energy eigenvalues at the TRIM points~\cite{Wang2012b}.

When the $z$ component $S_z$ of the total spin is conserved, the topological invariant 
can be obtained by directly calculating the spin Chern number expressed 
in terms of the single-particle Green's function. 
However, in $t_{2g}$ systems with the SOC, the spin Chern number is hardly 
formulated because the up and down spin sectors of the single-particle Green's
function always couple together.
The approach based on the topological Hamiltonian is best suited to the numerical 
calculation of the topological invariant for the interacting $t_{2g}$ systems.
Therefore, this method has been adopted very often to explore
the topological properties of many interacting 
systems~\cite{Yosida2013,Witczak2014,Hung2014,Gu2015,Grandi2015,He2016a}.
Moreover, it has been shown that the method is enough to obtain relevant 
results on the topological phase transition in interacting systems
as long as the electronic and topological phases can be certainly defined 
by fermionic degrees of freedom~\cite{He2016b}.

\section{Noninteracting system}\label{sec:non-int}

\subsection{Topological phase diagram}

\begin{figure*}[t]
\centering
\includegraphics[width=1.8\columnwidth]{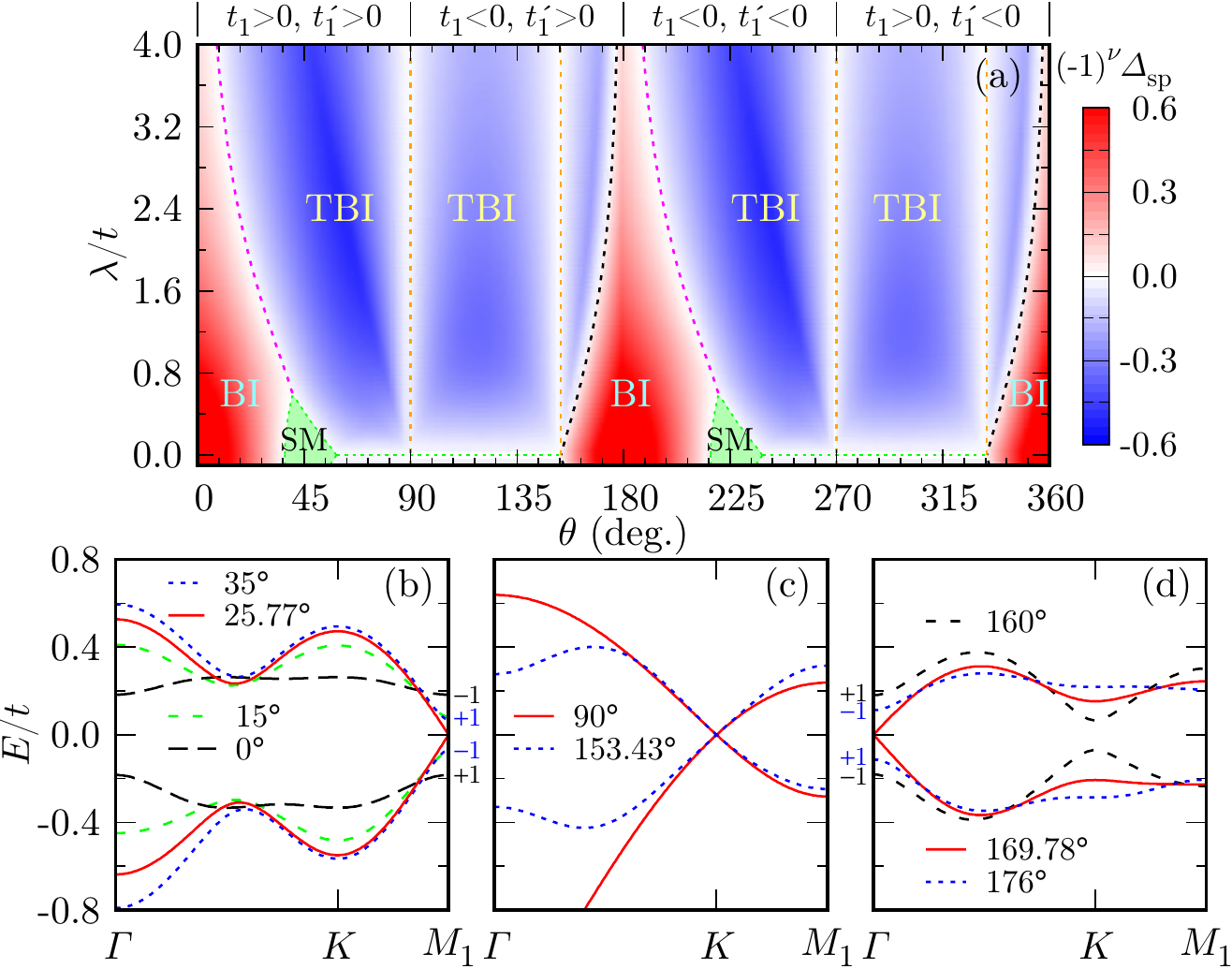}
\caption{
 (a) Topological phase diagram for the $t_{2g}^5$ system obtained from
 the product of the single-particle gap $\Delta_{\rm sp}\,(\ge0)$ 
 and the topological invariant $(-1)^\nu$ 
  with respect to the 1st NN hopping parameter
 $\theta$ (in $t_1=t \cos \theta$ and $t_1'=t \sin \theta$) and the spin-orbit 
 coupling $\lambda$.  
 BI and TBI denote trivial band insulator and topological band insulator 
 phases, respectively.
 A green region indicates a semimetallic (SM) phase.
 Dashed lines with black, magenta, and orange colors represent 
 regions where Dirac dispersions appear across the Fermi energy 
 at the $\Gamma$, $M$, and $K$ points, respectively, and thus $\Delta_{\rm sp}=0$. 
 A green dashed line at $\lambda=0$ represents the semimetal region with $\Delta_{\rm sp}=0$. 
 (b)--(d) Energy band dispersions around the Fermi energy ($E=0$)
 for various values of $\theta$ (indicated in the figures) when $\lambda=1.6t$.
 Dirac dispersions appear at the $M$ points when $\theta \approx 25.77^\circ$, 
 at the $K$ and $K'$ points when $\theta=90^\circ$ and $\theta \approx 153.43^\circ$,
 and at the $\Gamma$ point when $\theta \approx 169.78^\circ$.
 Parity eigenvalues for (b) $\theta=0^\circ$ and $35^\circ$ at the $M_1$ point and
 (d) $\theta=160^\circ$ and $176^\circ$ at the $\Gamma$ point are 
 also indicated with black and blue colors, respectively. 
 Here we assume that $t>0$ and set that $t_2=t_2'=t_3=\Delta_{\rm tr}=0$.
 The signs of $t_1$ and $t_1'$ are indicated on the top of (a). 
}
\label{Fig_NN}
\end{figure*}

First, we explore the role of the 1st NN hopping channel on the topological property.
Figure~\ref{Fig_NN}(a) shows the topological phase diagram as functions 
of the 1st NN hopping parameter $\theta$ and the SOC strength $\lambda$.
The phase diagram is obtained by calculating the product of 
a band gap $\Delta_{\rm sp}\,(\ge0)$ 
and a topological invariant $(-1)^\nu$ given in Eq.~(\ref{Eq_nu}). 
This quantity is exactly the same as a topological mass of the Kane-Mele model
when the band gap is determined at TRIM points~\cite{Murakami2007}. 
Red and blue regions in Fig.~\ref{Fig_NN}(a) correspond to 
topologically trivial BI and nontrivial $Z_2$ TBI phases, respectively. 
A semimetallic region, in which the highest energy of the valence bands
is larger than the lowest energy of the conduction bands,
is indicated by green in Fig.~\ref{Fig_NN}(a). 

Since the reversal of the hopping parameters (i.e., $\theta\to\theta+\pi$) does not change 
the energy band dispersions and the corresponding Bloch wave functions but only alter the 
sign of their parity eigenvalues, 
all topological quantities $\delta_{\Lambda_i}$ for $\theta+\pi$ 
have the opposite signs of those for $\theta$ 
when odd numbers of Kramers pairs are occupied. 
Note that there are 10 electrons per unit cell in our $t_{2g}^5$ system in the honeycomb lattice, 
and hence odd numbers of Kramers pairs are occupied at each TRIM point.  
However, this never changes the $Z_2$ topological invariant 
because there are four TRIM points in our system. 
Therefore, the topological phase diagram for $180^\circ \le \theta \le 360^\circ$
is exactly the same as that for $0^\circ \le \theta \le 180^\circ$
as shown in Fig.~\ref{Fig_NN}(a).

When $\theta$ is $0^\circ$ or $180^\circ$, 
$t_1'=0$ and only $t_1$ contributes to the hopping. In this limit, 
the electronic energy band structures without the SOC can be interpreted 
in terms of quasimolecular orbitals formed in each hexagon of the honeycomb lattice,  
which are well separated in energy and characterized by the parity eigenstates~\cite{Mazin2012,Foyevtsova2013}. 
Once the hopping $t_1$ is considered with the other terms kept absent, the six-fold degenerate $t_{2g}$ bands 
(two sites per unit cell without considering the spin degree of freedom) are split into dispersionless bands with 
$a_{1g}$, $e_{2u}$, $e_{1g}$, and $b_{1u}$ symmetries, charactering the quasimolecular orbitals, 
whose energies are $2t_1$, $t_1$, $-t_1$ and $-2t_1$, respectively. 
Therefore, the highest unoccupied band for the $t_{2g}^5$ configuration is the band with $a_{1g}$ symmetry 
for $t_1>0$ or $b_{1u}$ symmetry for $t_1<0$.
Based on the analytic form of the quasimolecular orbital with $a_{1g}$ ($b_{1u}$) symmetry  
in Ref.~\onlinecite{BHKim2016}, 
we can easily show that the parity eigenvalues at the $\Gamma$, $M_1$, $M_2$,
and $M_3$ points of the highest unoccupied band, i.e., $\xi_6(\Gamma)$, $\xi_6(M_1)$, $\xi_6(M_2)$, 
and $\xi_6(M_3)$, are $+1$, $-1$, $+1$, and $-1$ ($-1$, $+1$, $-1$, and $+1$), respectively.
Topological quantities $\left(\delta_{\Gamma},\delta_{M_1},\delta_{M_2},
\delta_{M_3}\right)$ are thus 
$(-1,+1,-1,+1)$ for $t_1>0$
and $(+1,-1,+1,-1)$ for $t_1<0$.
Therefore, topologically trivial insulator with $(-1)^\nu=+1$
is stabilized for both $\theta=0^\circ$ and $180^\circ$ without the SOC.
When the SOC increases, the energy band character smoothly changes from
the quasimolecular to relativistic $j_{\rm eff}$ bands~\cite{BHKim2016}.
However, no gap closure happens at the Fermi energy and  
the topological invariant remains the same regardless of the strength of $\lambda$.
This is why the $Z_2$ number is always zero near $\theta=0^\circ$
and $180^\circ$ in Fig.~\ref{Fig_NN}(a), although the trivial BI region apparently 
decreases with increasing $\lambda$.

\begin{table}[b]
\centering
\caption{
Topological quantities $\delta_\Gamma$, $\delta_{M_1}$, $\delta_{M_2}$, and
$\delta_{M_3}$, and topological invariant $(-1)^\nu$ for the $t_{2g}^5$ system 
in the honeycomb lattice 
with several representative values of $\theta$ parametrizing 
the 1st NN hopping parameters.  
Here we set that $\lambda=1.6t$ and $t_2=t_2'=t_3=\Delta_{\rm tr}=0$, 
assuming that $t>0$. 
}
\label{Table_Z2}
\begin{ruledtabular}
\begin{tabular} { c c c c c c}
 $\theta$ & $\delta_\Gamma$ & $\delta_{M_1}$ & 
 $\delta_{M_2}$ & $\delta_{M_3}$ & $(-1)^{\nu}$  \\
\hline
 $0^\circ$ & $-1$ & $+1$ & $-1$ & $+1$ & $+1$ \\
 $35^\circ$ & $-1$ & $-1$ & $+1$ & $-1$ & $-1$ \\
 $160^\circ$ & $-1$ & $-1$ & $+1$ & $-1$ & $-1$ \\
 $176^\circ$ & $+1$ & $-1$ & $+1$ & $-1$ & $+1$
\end{tabular}
\end{ruledtabular}
\end{table}

When $\theta$ is away from $\theta=0^\circ$ or $180^\circ$, 
the strength of $t_1'$ increases and modifies the electronic and topological
characteristics. 
A finite band gap at the $M$ points ($\Gamma$ point) gradually decreases 
but a direct gap at the $\Gamma$ ($M$) point increases reversely 
when $\theta$ increases (decreases) from $0^\circ$ ($180^\circ$). 
Eventually, the valence and conduction bands touch each other at the $M$ ($\Gamma$) 
point and the Dirac like dispersion appears around the Fermi energy. 
Concomitantly, the parity eigenvalues at the three $M$ points ($\Gamma$ point) 
of the highest occupied and lowest unoccupied bands
are reversed with further increasing (decreasing) $\theta$.
Accordingly, topological quantities 
$(\delta_{\Gamma},\delta_{M_1},\delta_{M_2},\delta_{M_3})$ 
are changed from $(-1,+1,-1,+1)$ [$(+1,-1,+1,-1)$]
to $(-1,-1,+1,-1)$, as shown in Table~\ref{Table_Z2}.
This is well illustrated in Figs.~\ref{Fig_NN}(b) and \ref{Fig_NN}(d).
In the case of $\lambda=1.6t$, the Dirac like dispersions appear at the three $M$ points ($\Gamma$ point)
when $\theta$ is about $25.77^\circ$ ($169.78^\circ$).
A dashed line with magenta (black) color in Fig.~\ref{Fig_NN}(a) represents 
the topological phase boundary where the Dirac like energy band dispersion with 
$\Delta_{\rm sp}=0$ appears at the $M$ points ($\Gamma$ point).
Thus, the increase (decrease) of $\theta$ gives rise to 
the topological phase transition from a trivial BI [$(-1)^\nu=+1$] to a 
TBI [$(-1)^\nu=-1$]. 
We should emphasize that the TBI phase is realized in a much broader parameter region,
in sharp contrast with 
the previous report based on an effective $j_{\rm eff}=1/2$ model
in which the $t_1$ contribution on the 1st NN hopping channel between $j_{\rm eff}=1/2$ 
orbitals exactly cancels out~\cite{Shitade2009}.

When only $t_1'$ is finite (i.e., $\theta=90^\circ$),
the hopping between one specific orbital is allowed in each type of hoppings, 
i.e., $d_{XY}$ orbital for $Z$ type, $d_{YZ}$ orbital for $X$ type, and $d_{ZX}$ orbital 
for $Y$ type of the 1st NN hopping channel (see Table~\ref{hopM}). 
Therefore, when the SOC is absent, 
each orbital participates to form the bonding and antibonding states with the 
same type of orbital on the nearest neighboring sites connected 
through $t_1'$ along the different hoping direction.
This brings about two six-fold-degenerate Bloch states 
(including doubly degenerate spin states) 
with momentum independent dispersions. 
When $t_1$ is turned on, the momentum dependence arises in the dispersions 
and the six-fold degeneracy is lifted in the entire momentum space expect for 
$t_1=-2t_1'$, corresponding to $\theta=\pi\!-\!\cos^{-1}(\frac{2}{\sqrt{5}})\approx 153.43^\circ$, 
where the six-fold degeneracy still remains at the $\Gamma$, $K$, and $K'$ 
points. 
When the SOC is turned on, 
the six-fold degenerate states at the $K$ and $K'$ points
are split into low-energy two-fold degenerate states
and high-energy four-fold degenerate states. 
As in the graphene band, the Fermi energy crosses 
the four-fold degenerate bands at the $K$ and $K'$ points [also see Fig.~\ref{Fig_NN}(c)]. 
Therefore, the zero gap region with $\Delta_{sp}=0$ 
appears at $\theta=90^\circ$ and 
$\theta=\pi\!-\!\cos^{-1}(\frac{2}{\sqrt{5}}) \approx 153.43^\circ$, 
regardless of $\lambda$ values, as indicated by orange dashed lines in  Fig.~\ref{Fig_NN}(a). 
Figure~\ref{Fig_NN}(c) shows the more detailed energy band structure at these $\theta$
values. Clear Dirac like dispersions appear at both $K$ and $K'$ points, while there is a finite gap at the 
$\Gamma$ point. 
Since both $K$ and $K'$ are not the TRIM points, 
the band gap closure at these points does not alter the parity 
eigenvalues of the occupied bands at the TRIM points.
Topological quantities
$(\delta_{\Gamma},\delta_{M_1},\delta_{M_2},\delta_{M_3})$ 
are always $(-1,-1,+1,-1)$ across these values of $\theta$
and thus nontrivial $Z_2$ topology is still robust.

The 1st NN hopping between the relativistic $j_{\rm eff}=1/2$ orbitals
is exactly cancelled when only $t_1$ is considered. 
The other hopping process attributed to $t_1'$ can give rise to a finite 1st NN hopping 
in the effective Kane-Mele model of the $j_{\rm eff}=1/2$ manifold. 
However, $t_1$ still leads to a finite 1st NN hopping between the $j_{\rm eff}=1/2$ 
and $3/2$ orbitals.
The virtual hopping process via $j_{\rm eff}=1/2 \to 3/2 \to1/2$ orbital is enough to
give rise to the effective hopping between the 2nd and 3rd NN sites in the 
$j_{\rm eff}=1/2$ manifold~\cite{Catuneanu2016}.
Thus, we expect that the variation of $\theta$ parameter in our $t_{2g}$ model 
induces the relative enhancement of the 2nd and 3rd NN hopping strengths in the effective Kane-Mele model. 
Therefore, the topological phase transition found here 
by varying relative
strengths of the two processes in the 1st NN hopping channel of our system can be understood
in the analogy of the Kane-Mele model with the 2nd and 3rd NN hopping channels.

Recently, Laubach {\it et al}. have reported 
a similar topological phase diagram of a $t_{2g}$ band model 
with respect to the relative strength of two 1st NN hopping processes
and the SOC~\cite{Laubach2017}.
The two 1st NN hopping processes considered in their model are those that lead to
the Kitaev-type and Heisenberg-type magnetic interactions 
in the strong coupling limit. 
The former is exactly the same as our $t_1$ hopping. 
The latter is the hopping processes with our $t'_1$ hopping and 
additional hoppings among the same orbitals. 
We consider the direct hopping only between, e.g., $d_{XY}$ orbitals in the $Z$ type,
whereas they consider the direct hopping between all $t_{2g}$ orbitals, including also, e.g., 
$d_{YZ}$ and $d_{ZX}$ orbitals in the $Z$ type. 
Although this difference in the hopping parameters gives rise to
an additional metallic region in a small SOC region around $\theta=90^\circ$ 
(not found here in our phase diagram), 
the topological phase diagram is in good agreement with our result for $0\le \theta \le \pi/2$
because their study is limited for both hopping processes positive. 
Therefore, the topological phase transition with the gap closure 
at the $\Gamma$ point appering in $\pi/2\le \theta \le \pi$ is not found in their study.

\subsection{Edge state}

\begin{figure*}[t]
\centering
\includegraphics[width=1.8\columnwidth]{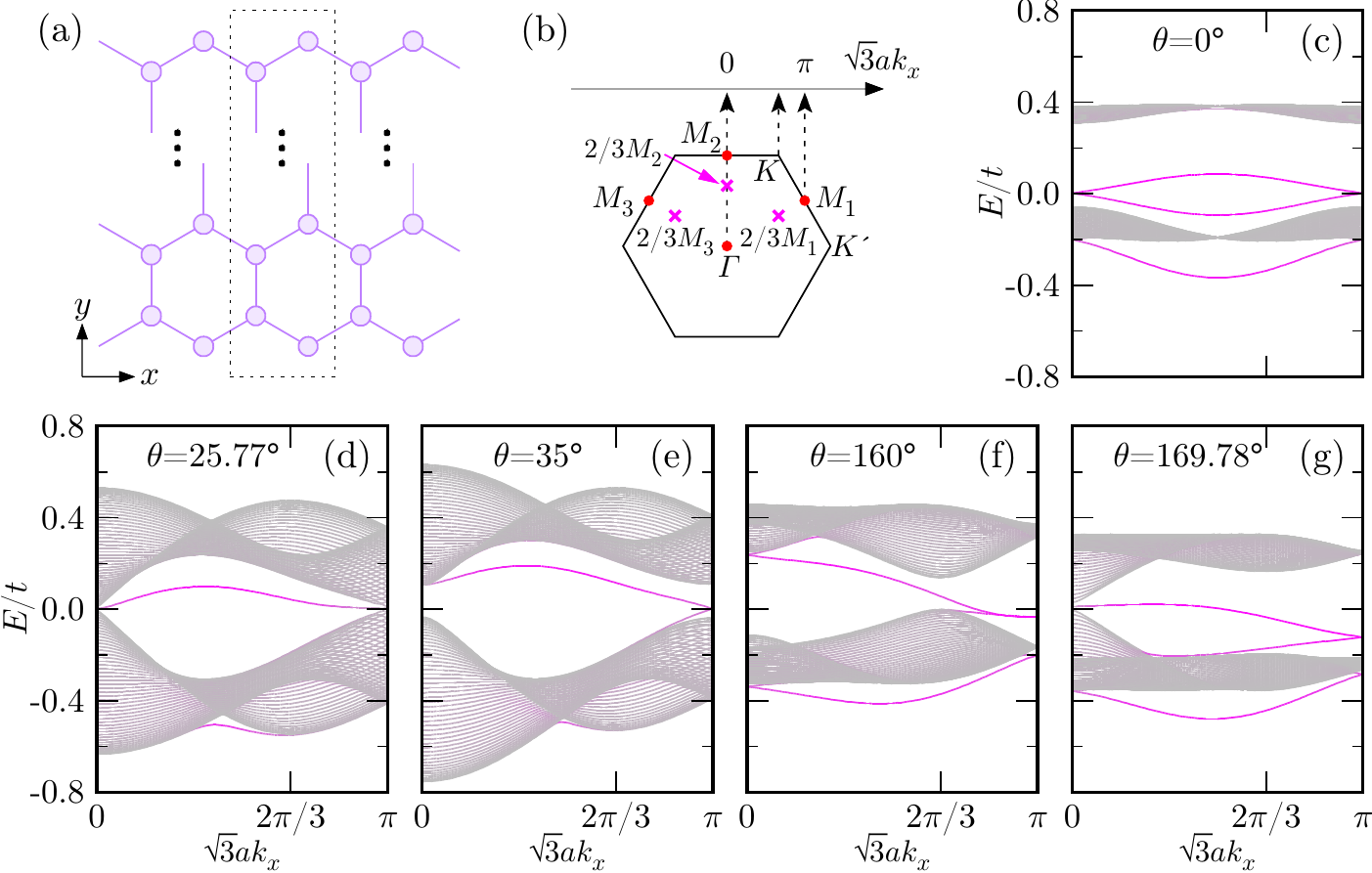}
\caption{
  (a) Schematic diagram of a zigzag stripy geometry of the honeycomb lattice. 
  The unit cell indicated by a dashed box contains one hundred sites. 
  (b) Mapping of the crystal momenta between the two-dimensional 
  honeycomb lattice and the one-dimensional stripy geometry in (a). 
  High symmetric momenta are denoted by $\Gamma$: $(0,0)$, 
  $K$: $(\frac{2\pi}{3\sqrt{3}a},\frac{2\pi}{3a})$, $K'$: $(\frac{4\pi}{3\sqrt{3}a},0)$, 
  $M_1$: $(\frac{\pi}{\sqrt{3}a},\frac{\pi}{3a})$, $M_2$: $(\frac{2\pi}{3a},0)$, 
  and $M_3$: $(-\frac{\pi}{\sqrt{3}a},\frac{\pi}{3a})$. Representative momenta 
  at $\mathbf{k}=\frac{2}{3}(\frac{\pi}{\sqrt{3}a},\frac{\pi}{3a})$, 
  $\frac{2}{3}(\frac{2\pi}{3a},0)$, and $\frac{2}{3}(-\frac{\pi}{\sqrt{3}a},\frac{\pi}{3a})$ 
  are also indicated as 
  $2/3M_1$, $2/3M_2$, and $2/3M_3$, respectively.
  Here, $a$ is the distance between the 1st NN sites of the honeycomb lattice. 
  (c)--(g) Electronic energy band dispersions of the zigzag stripy geometry 
  for various values of the 1st NN hopping parameter 
  $\theta$ indicated in the figures. 
  The energy band dispersions dominated at the edges are highlighted
  with magenta. The Fermi energy is located at $E=0$. 
   Here we set that $\lambda=1.6t$ and $t_2=t_2'=t_3=\Delta_{\rm tr}=0$. 
} 
\label{Fig_edge}
\end{figure*}

One of the characteristic features of $Z_2$ TIs is the presence of 
symmetrically protected edge states which intersect the Fermi energy
odd numbers of times.
To explore the surface electronic structures of our $t_{2g}^5$ system in the honeycomb lattice, 
here we consider a zigzag stripy geometry of the lattice structure along the $x$-direction 
with fifty lattice sites along the $y$-direction, thus containing one hundred sites in the unit cell, 
as schematically depicted in Fig.~\ref{Fig_edge}(a). 
Because the translation symmetry is broken along the $y$-direction, 
the two-dimensional momentum of the honeycomb lattice is projected onto the 
one-dimensional one shown in Fig.~\ref{Fig_edge}(b).

Figures~\ref{Fig_edge}(c)--\ref{Fig_edge}(g) show the electronic energy band structures of the zigzag
stripy geometry for various values of $\theta$ with $\lambda=1.6t$.
The electronic bands dominantly contributed from the edges are highlighted with magenta. 
Intriguingly, the energy band dispersions manifested inside the bulk band gap 
originate from the edge states for all the parameter region of $\theta$. 
Because the IS
as well as the TRS is still preserved even in the stripy geometry, 
the energy dispersions $\varepsilon_{k_xs}^{({\rm U})}$ and $\varepsilon_{k_xs}^{({\rm L})}$ 
($s=\,\Uparrow,\Downarrow$: pseudospin) of the edge states at upper and lower edges, 
respectively, are related as $\varepsilon_{k_x\Uparrow}^{(\rm U)}=\varepsilon_{-k_x\Uparrow}^{(\rm L)}
=\varepsilon_{k_x\Downarrow}^{(\rm L)}=\varepsilon_{-k_x\Downarrow}^{(\rm U)}$, where the first 
and third equalities are due to the inversion operation and the second equality is due to the time reversal operation. 
Therefore, the edge states show four-fold degeneracy 
at the TRIM points, i.e., $k_x=0$ and $\frac{\pi}{\sqrt{3}a}$
where $a$ is the distance between the 1st NN sites, 
irrespective of the width of the zigzag stripy geometry of the lattice.  
At any momentum away from these momenta, however, the surface bands are simply doubly degenerate. 
As show in Figs.~\ref{Fig_edge}(c)--\ref{Fig_edge}(g), these two-fold degenerate surface bands are eventually 
connected to other surface bands at $k_x$ and $\frac{\pi}{\sqrt{3}a}=0$ with quite different ways depending 
on the bulk topological feature.

As shown in Fig.~\ref{Fig_edge}(c), in the topologically trivial BI phase at and close to $\theta=0^\circ$ 
and $180^\circ$, 
the surface bands located inside the bulk band gap 
are well isolated from the bulk conduction and valence band continua and 
connect pairwise at the TRIM points. 
This is reminiscence of the energy dispersion at the edge of a single-layer Na$_2$IrO$_3$ 
recently studied by Catuneanu {\it et al.}~\cite{Catuneanu2016}.
With increasing or decreasing $\theta$ from $0$ or $180^\circ$, 
some part of the surface bands is buried in the valence band continuum 
but they never contact the conduction band continuum until the bulk band gap is closed
at the $\Gamma$ or $M$ points.
Thus, the surface bands clearly intersect the Fermi energy 
even number of times, as expected for a topologically trivial BI.

When the bulk gap is closed at the $M$ points for $\theta=25.77^\circ$ 
or at the $\Gamma$ point for $\theta=169.78^\circ$ [see Figs.~\ref{Fig_NN}(b) and \ref{Fig_NN}(d)], 
the bulk conduction and valence band continua touch each other 
$k_x=\frac{\pi}{\sqrt{3}a}$, as shown in Fig.~\ref{Fig_edge}(d), or at 
$k_x=0$, as shown in Fig.~\ref{Fig_edge}(g). 
With further increasing or decreasing $\theta$, the bulk conduction and valence band continua depart 
and the surface bands in the bulk band gap are again well separated from the bulk continua. However, the 
connectivity of the surface bands qualitatively changes.    
The pairwise connection of the surface bands is now broken and the 
surface bands cross the Fermi energy from the bulk conduction band continuum 
to the bulk valence band continuum, as shown in Figs.~\ref{Fig_edge}(e) and \ref{Fig_edge}(f).
Thus, the surface bands intersect the Fermi energy odd number of times, as expected in the TI phase.

\subsection{Further neighboring hopping and trigonal distortion}

According to previous studies, the electronic and topological properties 
of Na$_2$IrO$_3$ and its isostructural systems depend sensitively on the 
further neighboring hopping channels or the 
local electronic modulation induced by structural distortions.
Here we investigate the effects of the 2nd and 3rd NN hopping channels and 
the trigonal distortion on the topological phase diagram.

\begin{figure}[t]
\centering
\includegraphics[width=0.95\columnwidth]{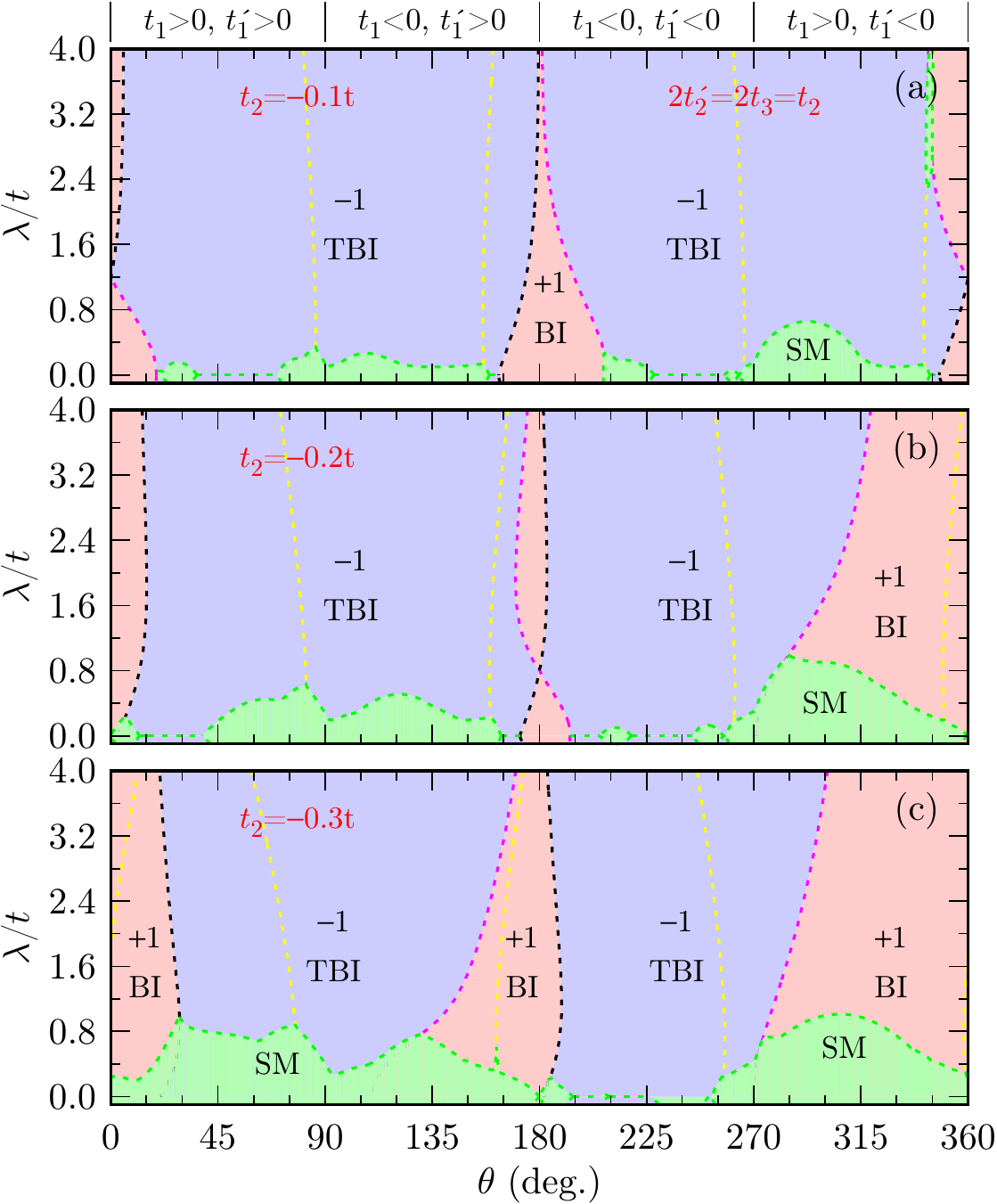}
\caption{
 Topological phase diagrams with respect to the 1st NN hopping 
 parameter $\theta$
 (in $t_1=t\cos\theta$ and $t_1'=t\sin\theta$) and the spin-orbit coupling $\lambda$
 for (a) $t_2=-0.1t$ and $t_2'=t_3=-0.05t$,
 (b) $t_2=-0.2t$ and $t_2'=t_3=-0.1t$, and 
 (c) $t_2=-0.3t$ and $t_2'=t_3=-0.15t$.
 Here we set $\Delta_{\rm tr}=0$. 
 Light red and blue areas represent topological band insulator (TBI) and trivial band insulator (BI) 
 phases with $(-1)^\nu=-1$ and $+1$, respectively.
 Semimetallic (SM) phases are stabilized in light green area.
 Black, magenta, and yellow dashed lines represent regions where Dirac dispersions appear
 at the $\Gamma$, $M$, and $K$ points, respectively, with the Dirac points located exactly at the Fermi energy.
 Green dashed lines at $\lambda=0$ represent the semimetallic regions 
 with $\Delta_{\rm sp}=0$. 
 The signs of $t_1$ and $t_1'$ are indicated on the top of (a). 
}
\label{Fig_NNN}
\end{figure}

Figure~\ref{Fig_NNN} shows the topological phase diagrams
for different 2nd and 3rd NN hopping parameters. 
Here we simply set $t_2=2t_3$ and $t_2'=t_3$ because 
this is not far from the theoretical estimations 
for Na$_2$IrO$_3$ and its isostructural materials 
(also see table~\ref{Table_EP}). 
TBI and BI phases are determined by
the topological invariant $(-1)^\nu$ in Eq.~(\ref{Eq_nu}). 
When $t_2$ varies from $0$ to $-0.3t$, 
the semimetal region is enlarged, and the BI-TBI phase boundaries 
indicated by black dashed lines, where the Dirac dispersion appears at $\Gamma$ point, 
shifts rightward, 
whereas the other phase boundaries indicated by magenta dashed lines, 
where the Dirac dispersions appear at the $M$points, 
shift oppositely. Therefore, 
comparing to the topological phase diagram shown in Fig.~\ref{Fig_NN}(a), 
the topological insulating region is slightly enlarged when the 2nd and 3rd NN hopping channels 
are introduced. However, with further increasing the 2nd and 3rd NN hopping strengths, 
the topological insulating region decreases and in particular the TBI phase is largely suppressed  
for $270^\circ \le \theta \le 360^\circ$ (i.e., $t_1>0$ and $t_1'<0$).

\begin{figure}[t]
\centering
\includegraphics[width=0.95\columnwidth]{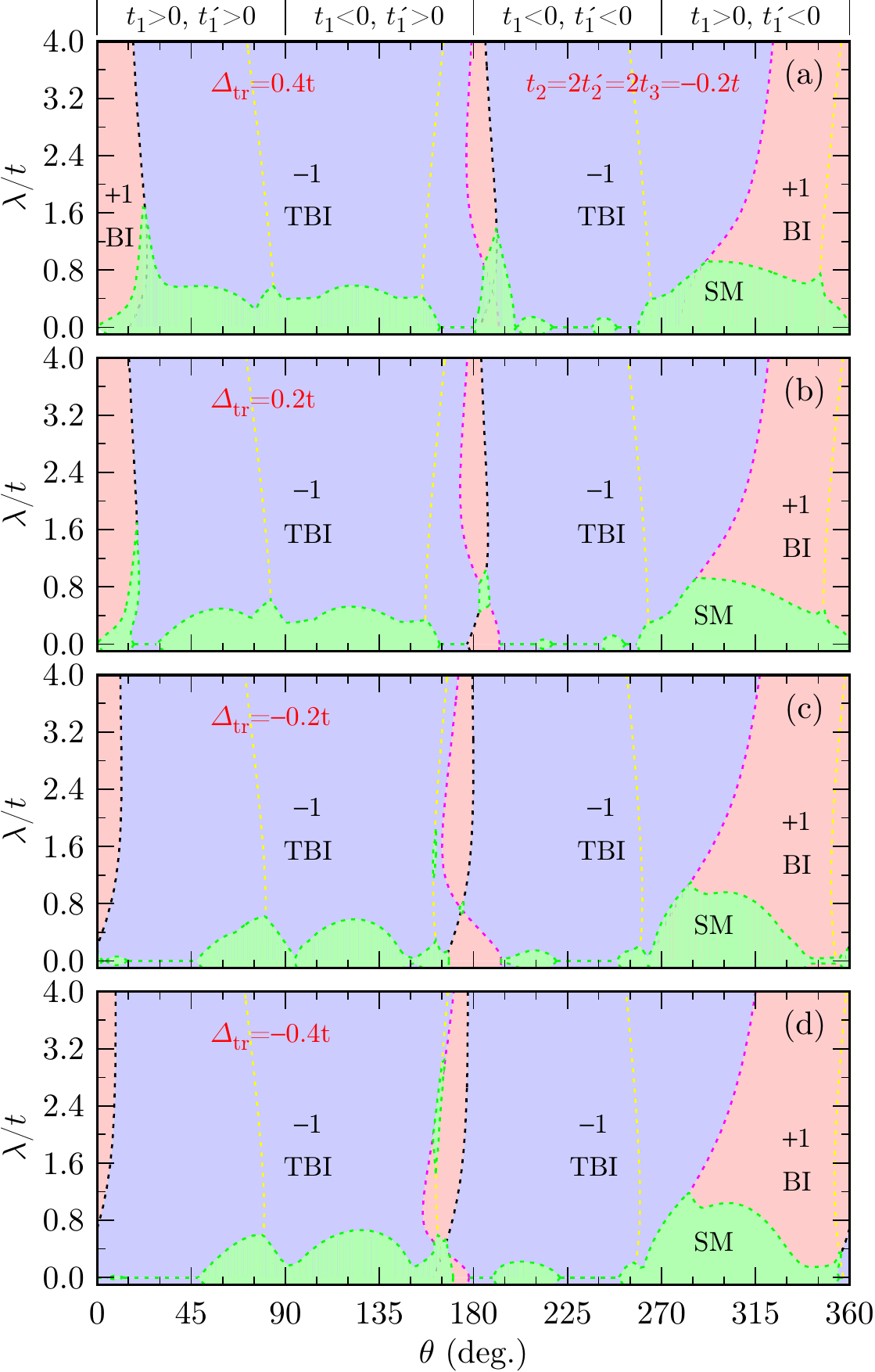}
\caption{
 Topological phase diagrams with respect to the 1st NN hopping parameter $\theta$
 (in $t_1=t\cos\theta$ and $t_1'=t\sin\theta$) and the spin-orbit coupling $\lambda$
  for (a) $\Delta_{\rm tr}=0.4t$, (b) $0.2t$, (c) $-0.2t$, and (d) $-0.4t$. 
 Here we set $t_2=-0.2t$ and $t_2'=t_3=-0.1t$ for the 2nd and 3rd NN hopping parameters. 
 Light red and blue areas represent topological band insulator (TBI) and trivial band insulator (BI) 
 phases with $(-1)^\nu=-1$ and $+1$, respectively.
 Semimetallic (SM) phases are stabilized in light green area.
 Black, magenta, and yellow dashed lines represent regions where Dirac dispersions appear
 at the $\Gamma$, $M$, and $K$ points, respectively, with the Dirac points located exactly at the Fermi energy.
 Green dashed lines at $\lambda=0$ represent the semimetallic regions 
 with $\Delta_{\rm sp}=0$. 
 The signs of $t_1$ and $t_1'$ are indicated on the top of (a). 
}
\label{Fig_TD}
\end{figure}

The trigonal distortion is also important to determine the topological phase.
Figure~\ref{Fig_TD} shows the topological phase diagrams for 
several values of the trigonal distortion $\Delta_{\rm tr}$ 
with $t_2=2t_2'=2t_3=-0.2t$ for the 2nd and 3rd NN hoppings.
The BI-TBI phase boundaries shift leftward
with decreasing $\Delta_{\rm tr}$ from positive to negative values.
Thus, the trigonal distortion affects the topological 
phases very differently depending on the relative strength of $t_1$ and $t_1'$. 
For instance, the topological phase at $\theta=15^\circ$ and $\lambda=1.6t$
changes from a trivial BI to a nontrivial TBI 
when $\Delta_{\rm tr}$ decreases from $0.4t$ to $-0.4t$.
In contrast, the topological phase at $\theta=300^\circ$ and $\lambda=1.6t$ 
transforms from a trivial BI to a nontrivial TBI 
when $\Delta_{\rm tr}$ increases 
from $-0.4t$ to $0.4t$.

Kim {\it et al.} have performed the first-principles calculations based on the density functional theory (DFT) 
to estimate $t_1\approx0.25$ eV and $t_1'\approx-0.5$ eV for Na$_2$IrO$_3$~\cite{CHKim2012}, 
which corresponds to $\theta=296.6^\circ$, as shown in Table~\ref{Table_EP}. 
According to our calculations in Fig.~\ref{Fig_TD}, 
the TBI phase easily appears at this value of $\theta$ when $\Delta_{\rm tr}$ is
positively large.
Indeed, they have concluded that a weak TBI phase can be realized in Na$_2$IrO$_3$ 
when there is the large trigonal distortion with positive $\Delta_{\rm tr}$.
However, other DFT based studies have estimated quite different values of the 1st NN hopping parameters. 
Their estimated values correspond to $\theta$ less than $10^\circ$, as summarized in Table~\ref{Table_EP}.
In these values of $\theta$, our results expect the TBI phase to be more favorable 
when $\Delta_{\rm tr}$ is negatively large, not positively large,  
as opposed to the prediction by Kim {\it et al.}~\cite{CHKim2012}.

\section{Interacting system}\label{sec:inter}

\begin{figure*}[t]
\centering
\includegraphics[width=1.6\columnwidth]{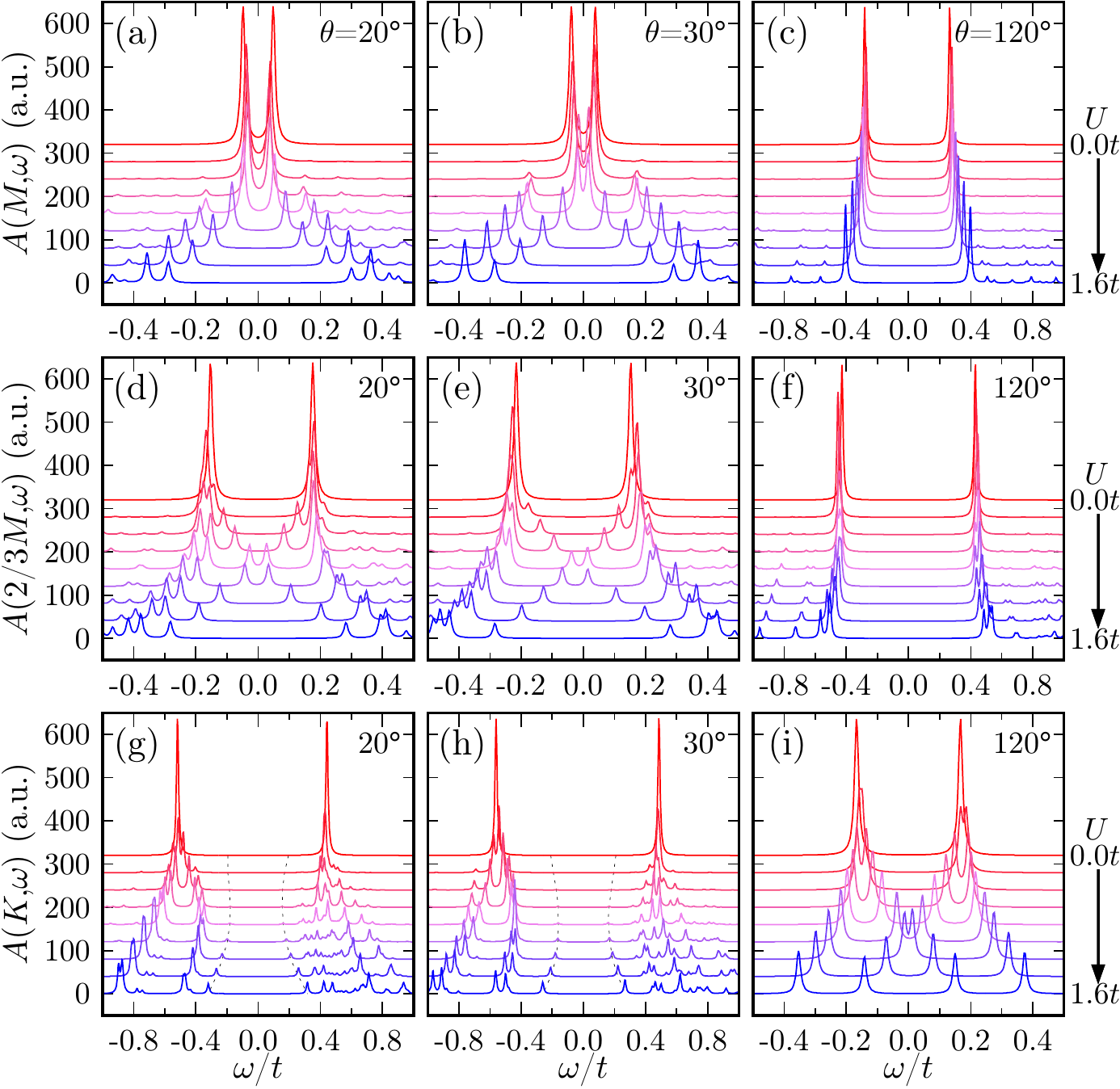}
\caption{
 Spectral function $A(\mathbf{k},\omega)$ of the single-particle Green's function 
 at the $M$, $2/3M$, and $K$ points with various $U$ values, indicated in the figures, 
 for (a), (d), and (g) $\theta=20^\circ$, (b), (e), and (h)  $\theta=30^\circ$,
 and (c), (f), and (i) $\theta=120^\circ$. 
 Notice that in each figure the spectral functions with different values of $U$ 
 are shifted from the top to the bottom in ascending order of $U$ with the increment of $0.2t$, for clarity.  
 We set that $\lambda=1.6t$ and $t_2=t_2'=t_3=\Delta_{\rm tr}=J_{\rm H}=0$. The Fermi energy is located at $\omega=0$. 
 Dashed lines in (g) and (h) indicate the location of the low-energy excitations showing weak-intensity 
 subpeak structures that determine the single-particle excitation gap at the $K$ point for $U\ne0$.
}
\label{Fig_SF}
\end{figure*}

To explore the effect of the Coulomb interaction on the electronic and
topological phases, here we consider a simple system with 
$\lambda=1.6t$ and $t_2=t_2'=t_3=\Delta_{\rm tr}=0$. 
First, we focus on the role of the on-site Coulomb repulsion $U$ 
in Eq.~(\ref{Eq:Corr}) by setting $J_{\rm H}=0$.
The CPT is employed to calculate the spectral function $A(\mathbf{k},\omega)$ of the 
single-particle Green's function [see Eq.~(\ref{Eq_SF}) for the definition] 
for various values of $\theta$ and $U$. 
Figure~\ref{Fig_SF} shows the representative results of the spectral function 
at the $M$, $2/3M$ [$2/3M_1$, $2/3M_2$, and $2/3M_3$ indicated by magenta crosses in Fig.~\ref{Fig_edge}(b)], 
and $K$ points 
as a function of $U$ for three different values of $\theta$ 
(i.e., $\theta=20^\circ$, $30^\circ$, and $120^\circ$). 
These three cases exhibit three different types of the electronic phase transition 
from a BI to a MI with increasing $U$: 
the phase transitions with the single-particle excitation gap closing at the $2/3M$ points, 
with the single-particle excitation gap closing consecutively at the $M$ and $2/3M$ points, 
and with the single-particle excitation gap closing at the $K$ and $K'$ points.

In the noninteracting limit, the spectral function is simply 
composed of the delta functions locating 
exactly at the energy of the noninteracting band dispersions. 
In finite $U$, the electron correlation induces 
the nonzero self-energy that generates additional peak structures in the 
spectral function. Because the electron coherency becomes poor due to 
the scattering among electrons, 
the spectral function becomes usually broader and the spectral weight can be 
even redistributed involving a large energy scale of $U$. 
These modifications of the spectral function certainly lead to the change of the single-particle 
excitation gap $\Delta_{\rm sp}$ determined by the two lowest excitations below and above 
the Fermi energy.

\begin{figure*}[t]
\centering
\includegraphics[width=2.\columnwidth]{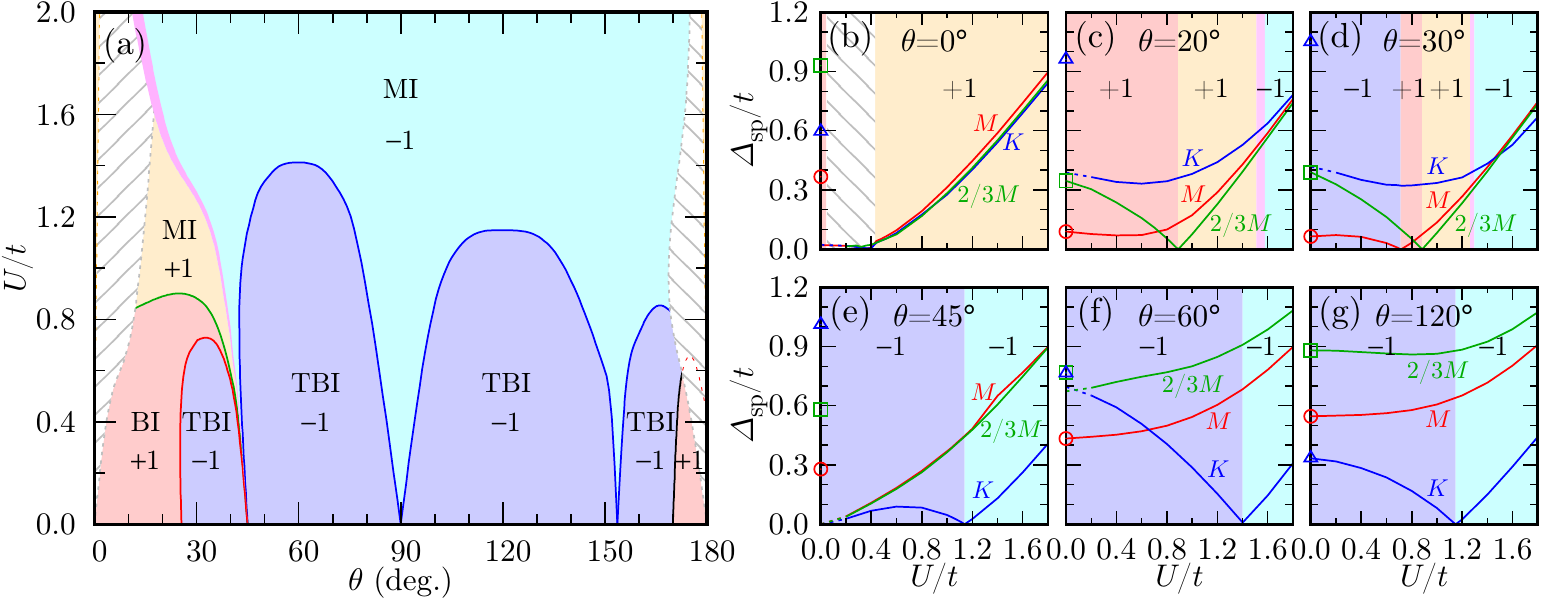}
\caption{
 (a) Topological phase diagram with respect to the 1st NN hopping parameter
 $\theta$ (in $t_1=t\cos\theta$ and $t_1'=t\sin\theta$) and the Coulomb repulsion $U$
 for $\lambda=1.6t$.
 Other parameters $\Delta_{\rm tr}$, $t_2$, $t_2'$, $t_3$, and $J_{\rm H}$ are set to be zero.
 Light blue (red) and cyan (orange) regions represent 
 the band and Mott insulator phases, respectively, with topological invariant 
 $(-1)^\nu=-1$ ($+1$).
 BI, TBI, and MI denote trivial band insulator phase, 
 topological band insulator phase, and Mott insulator phase, respectively.
 Black, red, green, and blue solid lines are phase boundaries 
 in which the single-particle excitation gap $\Delta_{\rm sp}$ is zero 
 at the $\Gamma$, $M$, $2/3M$, and $K$ points, respectively. 
 The topological characters in hashed regions are hard to be determined
 in our calculations. 
 (b)--(g) Single-particle excitation gap $\Delta_{\rm sp}$ at the $M$,
 $2/3M$, and $K$ points 
 as a function of $U$ for various values of $\theta$ ($0^\circ$, $20^\circ$,
 $30^\circ$, $45^\circ$, $60^\circ$, and $120^\circ$). 
 Open red circles, green squares, and blue triangles at $U=0$ indicate the direct excitation gaps 
 at the $M$, $2/3M$, and $K$ points, respectively, in the noninteracting system. 
 Notice that 
 the single-particle excitation gap for $\theta \le 60^\circ$ 
 does not necessarily approach asymptotically to the gap of the noninteracting system in the limit of $U\to0$, 
 as indicated by dotted lines near $U=0$ in (b)--(f). 
 This is because the weak-intensity subpeak structures appear inside the noninteracting gap for finite $U$,
 as shown in Figs.~\ref{Fig_SF}(g) and \ref{Fig_SF}(h).
}
\label{Fig_U}
\end{figure*}

The insulating gap $\Delta_{\rm sp}$ in the single-particle excitations for the MI is directly attributed to the Coulomb 
repulsion. It is easy to conjecture that 
$\Delta_{\rm sp}$ is monotonically increased with increasing $U$. 
In the BI, however, the insulating gap is already opened, without $U$, according to 
its own electronic kinetics. 
Because the Coulomb repulsion inhibits its kinetic effect,
the insulating gap $\Delta_{\rm sp}$ would be decreased and can be even diminished with increasing $U$.
Indeed, this feature has already been observed in our previous calculations of various spectroscopic quantities 
such as optical conductivity when $\theta=0^\circ$~\cite{BHKim2016}. 
As shown in Fig.~\ref{Fig_SF}, this is also the case 
in our systems studied here; the single-particle excitation gap in $A(\mathbf{k},\omega)$ 
at the $M$, $2/3M$, and/or $K$ points 
first decreases and then start to increase with increasing $U$ from the noninteracting limit. 
This implies that the insulating nature is changed from a BI to a MI with increasing $U$.

Figures~\ref{Fig_U}(b)--\ref{Fig_U}(g) show the $U$ dependence 
of the single-particle excitation gaps at the $M$, $2/3M$, 
and $K$ points, $\Delta_{\rm sp}(M)$, $\Delta_{\rm sp}(2/3M)$, and $\Delta_{\rm sp}(K)$, respectively, estimated 
from the spectral functions for six different values of $\theta$. 
One of $\Delta_{\rm sp}(M)$, $\Delta_{\rm sp}(2/3M)$, and $\Delta_{\rm sp}(K)$
becomes zero at the critical $U$ value. 
These critical values at which $\Delta_{\rm sp}(M)=0$,
$\Delta_{\rm sp}(2/3M)=0$, and $\Delta_{\rm sp}(K)=0$ 
are drawn with red, green, and blue solid lines, respectively, 
in the topological phase diagram shown in Fig.~\ref{Fig_U}(a). 
When $\theta$ is larger than $169.78^\circ$, the single-particle excitation gap
at the $\Gamma$ point, $\Delta_{\rm sp}(\Gamma)$, can also be zero with increasing $U$. 
The corresponding critical $U$ values are indicated with black solid line in Fig.~\ref{Fig_U}(a). 

As shown in Figs.~\ref{Fig_SF}(g) and \ref{Fig_SF}(h), 
the two $\delta$-function peaks in the spectral function at the $K$ and $K'$ points 
near the Fermi energy for $U=0$ 
are split into multiple subpeaks as soon as finite $U$ is introduced. 
The two subpeaks closest to the Fermi energy, which determine the single-particle excitation 
gap for finite $U$, emerge at the energies rather away from the $\delta$-function 
peaks in the noninteracting system, as indicated by dashed lines 
in Figs.~\ref{Fig_SF}(g) and \ref{Fig_SF}(h). 
Their spectral weights gradually decreases with decreasing $U$ and completely vanishes at $U=0$. 
Therefore, the single-particle excitation gap does not necessarily approaches to that 
of the noninteracting system in the limit of $U\to0$, as indicated by dotted line 
near $U=0$ in Figs.~\ref{Fig_U}(b)--\ref{Fig_U}(f).

\begin{figure}[t]
\centering
\includegraphics[width=0.98\columnwidth]{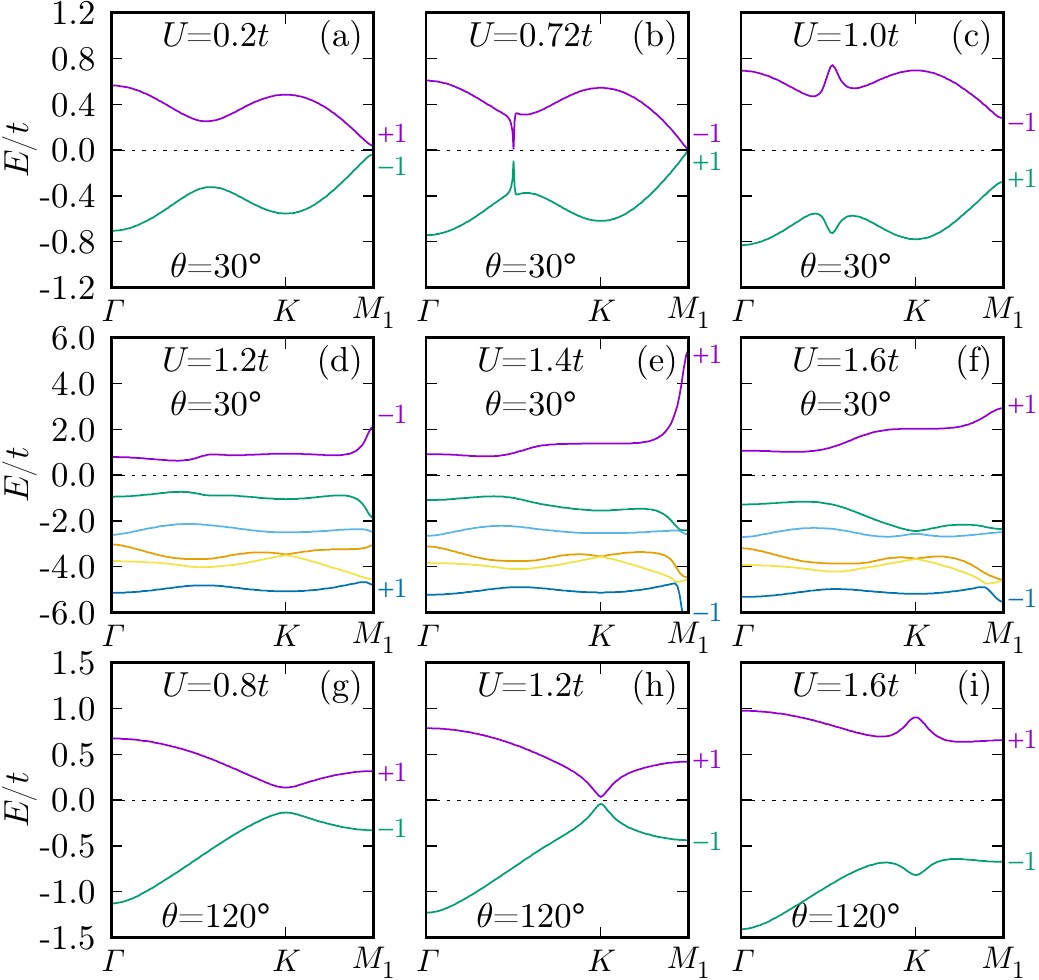}
\caption{
 Energy dispersions of the topological Hamiltonian $\mathbf{H}_{\rm T}(\mathbf{k})$ 
 for various $U$ and $\theta$ values indicated in the figures. 
 We set that $\lambda=1.6t$ and $t_2=t_2'=t_3=\Delta_{\rm tr}=J_{\rm H}=0$. 
 The Fermi energy is located at $E=0$. 
 ``$+1$" and ``$-1$" given above the Fermi energy 
 refer to the parity eigenvalue of the lowest conduction energy band at the $M_1$ point, 
 whereas those given blew the Fermi energy refer to the parity eigenvalue of the 
 highest [(a)--(c),(g)--(i)] or lowest valence energy band [(d)--(f)] at the $M_1$ point.
}
\label{Fig_TH}
\end{figure}

To explore the topological feature for finite $U$, 
we calculate the topological Hamiltonian $\mathbf{H}_{\rm T}(\mathbf{k})$ 
based on the CPT and evaluate the topological invariant $(-1)^\nu$ 
for the eigenstates of $\mathbf{H}_{\rm T}(\mathbf{k})$.
Figure~\ref{Fig_TH} shows examples of the energy dispersions of 
$\mathbf{H}_{\rm T}(\mathbf{k})$ for various $\theta$ and $U$ values.
Although the topological Hamiltonian $\mathbf{H}_{\rm T}(\mathbf{k})$ 
can mimic the topological properties of the interacting system perfectly,
the energy dispersion of $\mathbf{H}_{\rm T}(\mathbf{k})$ has no reason 
to be the same as that of the corresponding interacting
system because the energy dispersion for the latter is determined by the spectral function 
$A(\mathbf{k},\omega)$ of the 
single-particle Green's function. 
Only in a weakly interacting system, such as $U=0.2t$ in 
Fig.~\ref{Fig_TH}(a), where 
the electronic self-energy is almost zero, both dispersions are expected 
to be almost the same.
However, when $U$ is large, these two dispersions are evidently distinct, 
as shown in Fig.~\ref{Fig_TH} and Fig.~\ref{Fig_SFA}.
This has also been commonly observed in previous studies~\cite{Grandi2015,Gu2015}.

When the single-particle excitation gap in an interacting system is closed 
at a specific momentum $\mathbf{k}^*$, the spectral function $A(\mathbf{k}=\mathbf{k}^*,\omega)$
exhibits dominant spectral weight at the Fermi energy ($\omega=0$). 
In other words,  
$\mathbf{G}(\zeta,\mathbf{k})$ has poles at $\zeta=0$ and $\mathbf{k}=\mathbf{k}^*$.
Thus, the topological Hamiltonian, which is
proportional to $\mathbf{G}(0,\mathbf{k})^{-1}$,
should also exhibit the gap closure at the same momentum, simultaneously.
As shown in Figs.~\ref{Fig_TH}(b) and \ref{Fig_TH}(h) and Figs.~\ref{Fig_SFA}(b) and \ref{Fig_SFA}(h),
our calculation clearly manifests that 
the lowest conduction energy band and the highest valence energy band of the topological Hamiltonian 
touch the Fermi energy simultaneously at the critical $U$ value 
and at the momentum where the single-particle excitation gap of the interacting system is closed in the spectral function.

As the energy gap of $\mathbf{H}_{\rm T}(\mathbf{k})$ is closed at a TRIM point $\Lambda_i$, 
the parity eigenvalues of the corresponding conduction and valence energy bands of $\mathbf{H}_{\rm T}(\mathbf{k})$ 
at $\Lambda_i$ are exchanged. 
Accordingly, the topological quantities $\delta_{\Lambda_i}$ in Eq.~(\ref{Eq_nu}) is
reversed. 
If the gap closure happens at odd numbers of TRIM points,
the topological invariant $(-1)^\nu$ is reversed.
Therefore, in this case, the topological phase transition occurs. 
In the case of $\theta=30^\circ$, for instance, $\Delta_{\rm sp}(M)$ is closed 
at $U\approx0.71t$, as shown in Fig.~\ref{Fig_U}(c). 
Concomitantly, the parity eigenvalues of the lowest conduction and highest valence 
energy bands of $\mathbf{H}_{\rm T}(\mathbf{k})$ are reversed at the three $M$ points. 
Thus, the topological quantities $\left(\delta_{\Gamma},\delta_{M_1},\delta_{M_2},
\delta_{M_3}\right)$ of $\mathbf{H}_{\rm T}(\mathbf{k})$ change from
$\left(-1,-1,+1,-1\right)$ to $\left(-1,+1,-1,+1\right)$
when $U$ increases from below to above $U\approx0.71t$. 
This is an example where the Coulomb repulsion compels
the topological invariant $(-1)^\nu$ to change from $-1$ to $+1$,
hence representing the topological phase transition from the TBI to the trivial BI. 
Red and black solid lines in Fig.~\ref{Fig_U}(a) represent 
the parameters where the single-particle excitation gap in the spectral function is closed at 
the TRIM points, i.e., at the $M$ and $\Gamma$ points, respectively. 
Across these boundaries, the topological invariant $(-1)^\nu$ of $\mathbf{H}_{\rm T}(\mathbf{k})$ 
changes the sign between $-1$ and $+1$.

In the noninteracting case, the $Z_2$ topological invariant can be changed 
only when the single-particle excitation gap is closed. Therefore, 
the noninteracting single-particle Green's function has necessarily a pole at the Fermi energy 
exactly when the topological phase transition occurs. 
In the interacting case, however, this gap closure criteria is no longer 
mandatory.
In the presence of the interaction, the single-particle Green's function could have 
zeros along the real axis as well as poles~\cite{Dzyaloshinskii2003,Eder2008,Gurarie2011,Ezawa2013,Seki2017}. 
If the single-particle Green's function at a TRIM point $\Lambda_i$ becomes zero, instead of having a pole, 
at $\zeta=0$ when the topological phase transition occurs, 
the lowest conduction and highest valence energy bands of $\mathbf{H}_{\rm T}(\mathbf{k})$ 
do not touch each other at the Fermi energy. 
Instead, they are positively and negatively diverged, respectively, at $\Lambda_i$. 
Moreover, the parity eigenvalues of these diverging eigenstates of $\mathbf{H}_{\rm T}(\mathbf{k})$ 
are able to be exchanged at $\Lambda_i$ and thus the topological invariant can
be varied.

This is indeed observed in our calculations. 
As shown in Figs.~\ref{Fig_TH}(d) and \ref{Fig_TH}(e), 
the parity eigenvalues of the eigenstates of 
$\mathbf{H}_{\rm T}(\mathbf{k})$ with the largest and smallest eigenvalues 
are reversed at three $M$ points after their eigenvalues are diverged positively and negatively, respectively. 
Accordingly, the topological quantities 
$\left(\delta_{\Gamma},\delta_{M_1},\delta_{M_2},\delta_{M_3}\right)$ of 
$\mathbf{H}_{\rm T}(\mathbf{k})$ changes 
from $\left(-1,+1,-1,+1\right)$ to $\left(-1,-1,+1,-1\right)$,
keeping a finite single-particle excitation gap $\Delta_{\rm sp}(M)$ in the spectral function 
of the interacting system [see Figs.~\ref{Fig_SFA}(d) and ~\ref{Fig_SFA}(e)]. 
This type of topological phase transition is indicated with
light magenta lines in Figs.~\ref{Fig_U}(a), \ref{Fig_U}(c), and \ref{Fig_U}(d).

As shown in Figs.~\ref{Fig_U}(e)--\ref{Fig_U}(g), in the cases of $\theta=45^\circ$, $60^\circ$, and $120^\circ$, 
the single-particle excitation gap $\Delta_{\rm sp}(K)$ decreases first, 
diminishes at a certain $U$, and 
then increases with increasing $U$. 
Because the $K$ point is not a TRIM, 
the parity eigenvalues of $\mathbf{H}_{\rm T}(\mathbf{k})$ at the TRIM points 
remain the same even after this gap closure happens.  
Thus, the topological property of a MI region, indicated by light cyan color
in Fig.~\ref{Fig_U}, is the same as that of the noninteracting system with
$25.77^\circ \lesssim \theta \lesssim 169.78^\circ$, i.e., the QSH state.
Our calculations therefore affirm the possibility of the paramagnetic 
MI with nontrivial band topology in $t_{2g}^5$ honeycomb systems.

We should note here that we fail to calculate the topological invariant $(-1)^\nu$ of $H_{\rm T}(\mathbf{K})$ 
in parameter regions indicated by hatched areas in Figs.~\ref{Fig_U}(a) and \ref{Fig_U}(b). This is simply 
because $H_{\rm T}(\mathbf{K})$ evaluated by the CPT is broken down. More details are discussed in  
Appendix~\ref{app:th}.

\begin{figure}[t]
\centering
\includegraphics[width=0.95\columnwidth]{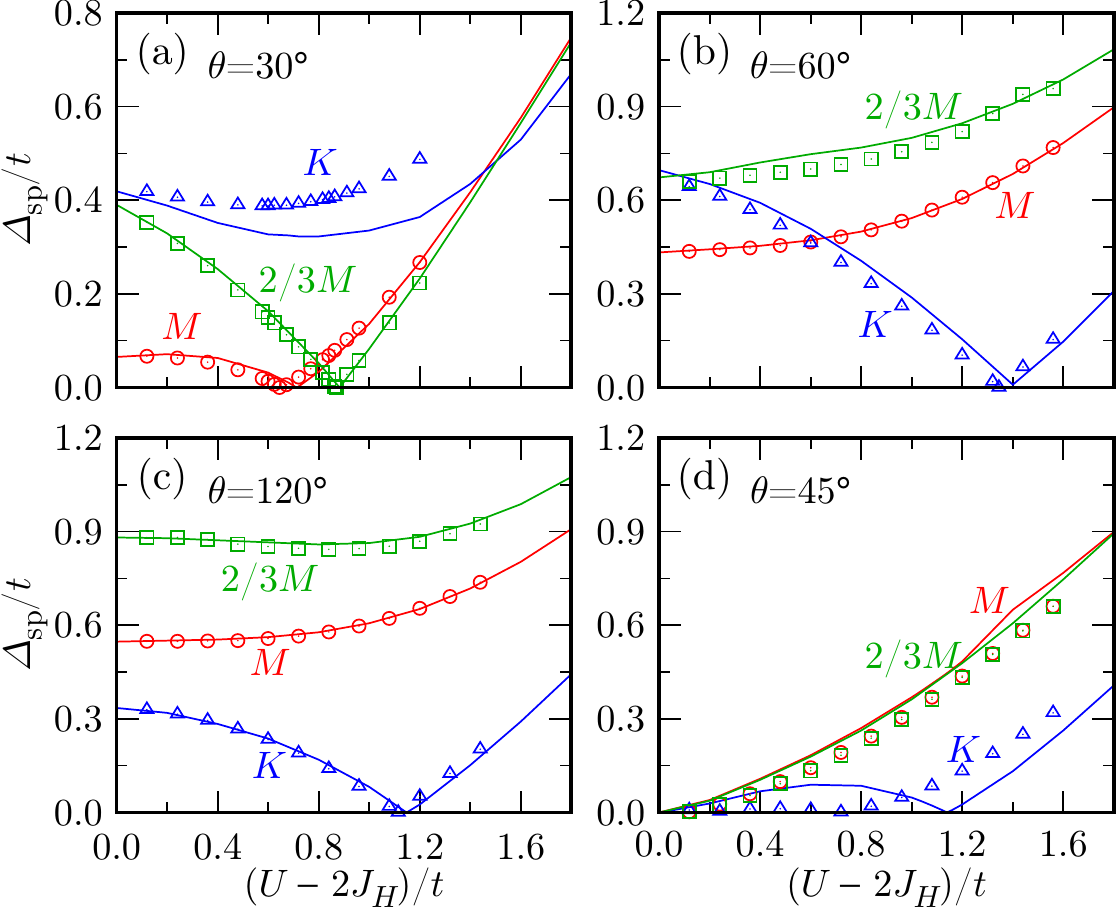}
\caption{
 Single-particle excitation gap $\Delta_{\rm sp}$ at the $M$ (red circles),
 $2/3M$ (green squares), and $K$ (blue triangles) points 
 as a function of $U$ for (a) $\theta=30^\circ$, (b) $60^\circ$,
 (c) $120^\circ$, and (d) $45^\circ$ when $J_{\rm H}=0.2U$.
 For comparison, the results for $J_{\rm H}=0$ are also shown by 
 red, green, and blue lines, corresponding to $\Delta_{\rm sp}$ 
 at the $M$, $2/3M$, and $K$ points, respectively. 
 We set that $\lambda=1.6t$ and $t_2=t_2'=t_3=\Delta_{\rm tr}=0$
}
\label{Fig_JH}
\end{figure}

It is often the case that the Hund's coupling $J_{\rm H}$ as well as the on-site Coulomb repulsion $U$ 
plays a crucial role in determining the electronic and magnetic properties 
of $t_{2g}$ systems.
For example, anisotropic magnetic exchange interactions such as Kitaev interaction 
are induced in the strong coupling limit of 
$t_{2g}^5$ systems with a honeycomb lattice structure 
only when $J_{\rm H}$ is finite~\cite{Chaloupka2010}. 
We thus investigate the effect of $J_{\rm H}$ on the electronic and topological phase diagram 
in a relatively weak coupling region.
Figure~\ref{Fig_JH} shows the single-particle excitation gap $\Delta_{\rm sp}$ 
for $\theta=30^\circ$, $45^\circ$, $60^\circ$, and $120^\circ$ when 
$J_{\rm H}=0$ and $0.2U$. 
Remarkably, we find that $\Delta_{\rm sp}$ for various $\theta$ values is
approximately scaled with $U-2J_{\rm H}$, 
which is the effective Coulomb interaction among different orbitals,
regardless of $J_{\rm H}$ values. 
As shown in Fig.~\ref{Fig_JH}(d), only $\Delta_{\rm sp}$ at the $K$ point near $45^\circ$ 
seems to deviate from this scaling. 
Therefore, we can conclude that the dominant effect of $J_{\rm H}$ 
on the electronic and topological phase diagram 
is the renormalization of the on-site Coulomb repulsion $U$.

\section{Discussion}\label{sec:discussion}

Table~\ref{Table_EP} summarizes the hopping and trigonal distortion parameters for 
the existing materials Na$_2$IrO$_3$, Li$_3$IrO$_3$, and $\alpha$-RuCl$_3$, 
which are extracted from literature~\cite{CHKim2012,HSKim2013,Foyevtsova2013,Yamaji2014,Winter2016}. 
In the case of Na$_2$IrO$_3$, the expected value of $\theta$ is about 
$5 \sim 10^\circ$~\cite{Foyevtsova2013,Yamaji2014,Winter2016}, 
although Ref.~\onlinecite{CHKim2012} reports that $\theta$ can be as large as $296^\circ$. 
Note that the SOC $\lambda$ for Na$_2$IrO$_3$ is $0.4 \sim 0.5$ eV and thus 
$\lambda/t$ is about $1.5 \sim 1.9 t$. 
Therefore, according to our results shown in Fig.~\ref{Fig_NNN}, 
this material can be in the TBI phase 
only when the 2nd and 3rd NN hopping strengths are within the proper range. 
In addition, as shown in Fig.~\ref{Fig_TD}, the phase boundary separating
the BI and TBI phases tends to shift leftward
as $\Delta_{\rm tr}/t$ decreases to be negative.
This infers that
strong negative $\Delta_{\rm tr}/t$ is more profitable 
for Na$_2$IrO$_3$ to be in the TBI phase.
However, the expected value of $t_2$ is about $-0.28 \sim -0.30 t$ and 
$\Delta_{\rm tr}/t$ is positive. 
These are pessimistic indications for Na$_2$IrO$_3$ being a TBI.

\begin{table}[b]
\centering
\caption{
The hopping and trigonal distortion parameters for Na$_2$IrO$_3$ 
and its isostructural Li$_2$IrO$_3$ and $\alpha$-RuCl$_3$, extracted from literature.
Because the atomic structure of these materials is slightly deviated from the ideal 
honeycomb lattice, we adopt the parameters from one among the three different types of 
the 1st and 2nd NN hopping channels. 
}
\label{Table_EP}
\begin{ruledtabular}
\begin{tabular} { c c c c c c c}
 & $t$ (meV) & $\theta$ ($^\circ$) & $t_2/t$ & $t_2'/t$ & $t_3/t$ &
 $\Delta_{\rm tr}/t$ \\
\hline
 Na$_2$IrO$_3$~\cite{CHKim2012} & 559.0 &296.6 & $-0.134$ & 
  $-0.134$ & $-0.134$ & $0.358$\\
 Na$_2$IrO$_3$~\cite{Foyevtsova2013} & 273.8 &10.0 & $-0.276$ & 
  $-0.133$ & - &  $0.085$\\
 Na$_2$IrO$_3$~\cite{Yamaji2014} & 276.4 & 8.9 & $-0.308$ & 
  $-0.109$ &  $-0.134$ & $0.101$\\
 Na$_2$IrO$_3$~\cite{Winter2016} & 264.3 & 5.7 & $-0.285$ & 
  $-0.137$ & $-0.133$ &  $0.086$ \\
 Li$_2$IrO$_3$~\cite{Winter2016} & 280.4 & 321.4 & $-0.203$ & 
  $-0.085$ & $-0.143$ & $0.134$\\
 $\alpha$-RuCl$_3$~\cite{HSKim2013} & 255.8 & 296.5 & $-0.227$ &
  $-0.078$ & $-0.192$ & - \\
 $\alpha$-RuCl$_3$~\cite{Winter2016} & 220.8 & 315.8 & $-0.268$ & 
  $-0.149$ & $-0.188$ & $0.090$
\end{tabular}
\end{ruledtabular}
\end{table}

Recently, Catuneanu {\it et al}. have studied theoretically the edge state of 
single-layer Na$_2$IrO$_3$~\cite{Catuneanu2016}. 
In their study, the effective Hamiltonian for the $j_{\rm eff}=1/2$ 
manifolds was constructed with the 1st, 2nd, and 3rd NN hoppings 
between $j_{\rm eff}=1/2$ orbitals, which are extracted from
the first-principles electronic band structure calculations of single-layer Na$_2$IrO$_3$.
The edge dispersion in the zigzag geometry was also calculated 
with the effective $j_{\rm eff}=1/2$ Hamiltonian. 
Their results are similar to the edge dispersion shown in Fig.~\ref{Fig_edge}(c).
This also confirms that Na$_2$IrO$_3$ is not in the TBI phase.

However, recent photoemission spectroscopy measurement on Na$_2$IrO$_3$ 
has observed a metallic band near the $\Gamma$ point~\cite{Alidoust2016,Moreschini2017}. 
If this metallic band is attributed dominantly to 
the surface honeycomb layer, the physical parameters in the surface honeycomb layer 
would be located very close to the topological phase boundary
because, according to our calculation in Fig.~\ref{Fig_NNN} and Fig.~\ref{Fig_TD},
the $\Gamma$-point Dirac dispersion appears at the Fermi energy in the phase boundary between 
the BI and TBI phases. 
This is an optimistic clue for the surface layer of Na$_2$IrO$_3$ to
be located not far from the TBI phase. 
Therefore, we expect that small structural tuning on Na$_2$IrO$_3$ would be enough to 
bring about the topological phase transition in the surface layer.

As shown in Table~\ref{Table_EP}, estimated $\theta$ for Li$_2$IrO$_3$ 
is about  320$^\circ$ and 
$\theta$ for $\alpha$-RuCl$_3$ is about $295^\circ$--$315^\circ$.
Therefore, according to the topological phase diagrams shown in Fig.~\ref{Fig_NNN} and Fig.~\ref{Fig_TD},
these materials could be in the TBI phase when $\Delta_{\rm tr}/t$ is
positively large, 
and the 2nd and 3rd NN hopping strengths are small enough~\cite{CHKim2012}. 
However, the estimated 2nd NN hopping strengths listed in Table~\ref{Table_EP} 
are relatively large ($|t_2|>0.2t$).
This implies that the TBI phase is hard to be stabilized in these parameters for Li$_2$IrO$_3$ and $\alpha$-RuCl$_3$.
For the realization of TBIs in these materials, 
it is advantageous to reduce the further neighboring hoppings.

Recently, Yamada {\it et al.} have proposed a new efficient way to experimentally control 
the hopping strengths of a honeycomb lattice
by introducing oxalate- or tetraaminopyrazine-based molecular ligands, 
instead of chlorine atoms, which are connected to the adjacent TM Ru$^{3+}$ ions~\cite{Yamada2016}. 
They have theoretically suggested that the relative strength of the 1st NN hopping channel 
can be tuned by selecting the molecular ligand. 
Since molecular ligands can also increase the distance between the adjacent TM ions
in a honeycomb lattice, it would be enough to modify 
the relative hopping strength of the 2nd and 3rd NN hoping channels.
Although their theory expects that these systems should be in the MI phase with 
the magnetic exchange interaction between the TM ions being properly designed, 
the topological phase could also be turned in the paramagnetic insulating limit.

In Sec.~\ref{sec:inter}, we have shown the possibility of 
the MI with nontrivial band topology
in a $t_{2g}^5$ system with the honeycomb lattice structure. 
However, the estimated $U$ values for 
Na$_2$IrO$_3$, Li$_2$IrO$_3$, and $\alpha$-RuCl$_3$ 
are much larger than $2t$ studied in Fig.~\ref{Fig_U}. 
Their $U-3J_{\rm H}$ values are estimated around 
$3.2t$ for Na$_2$IrO$_3$ and $6t$ for $\alpha$-RuCl$_3$~\cite{BHKim2016}. 
In such a large $U$ limit, the effective spin model with relativistic $J_{\rm eff}=1/2$ doublets is
expected to be a better description for the insulating state. 
 In the system with the honeycomb lattice structure, the magnetic exchange 
interaction between the 1st NN sites can be expressed with three different parameters: 
isotropic Heisenberg term ($J$), Kitaev term ($K$), 
and symmetric off-diagonal term ($\Gamma$)~\cite{Rau2014}. 
Furthermore, the relative strength of these three magnetic interaction terms can be 
varied with $\theta$. 
When $\theta=0^\circ$ and $180^\circ$, 
only the Kitaev term is accessible for finite Hund's coupling $J_{\rm H}$~\cite{Chaloupka2010}. 
Thus, in this case, the magnetic $Z_2$ spin liquid can be stabilized. 
In contrast, the Kitaev term is diminished and only the Heisenberg
term is survived when $\theta=90^\circ$. 
Thus, it gives rise to the antiferromagnetic Ne\'el order~\cite{Chaloupka2010}. 
Because the off-diagonal $\Gamma$ parameter is proportional to $t_1t_1'$,
its magnitude is maximum at $\theta=45^\circ$ and $135^\circ$, 
whereas it is absent at $\theta=0^\circ$, $90^\circ$, and $180^\circ$.

\section{Conclusion}\label{sec:conclusion}

We have investigated the topological property of a $t_{2g}^5$ system with 
a honeycomb lattice structure such as Na$_2$IrO$_3$ and 
the isostructural Li$_2$IrO$_3$ and $\alpha$-RuCl$_3$. 
By calculating the bulk topological invariant and the energy band dispersions of
edge states, we have unraveled that the hopping parameter $\theta$, which determines
the relative strength of the two different processes in the 1st NN hopping channel, plays 
an essential role in the topological phase transition between the trivial BI and the TBI. 
When the $pd\pi$-type hopping process mediated by
the edge-shared ligands is dominant, the topologically trivial phase
is favorable. On the other hand, when the $dd\sigma$-type direct hopping process becomes stronger, 
the topological phase transition occurs to the TBI phase at the critical $\theta$ 
where the band gap is closed at the $\Gamma$ or $M$ points.

We have also explored the topological phase transition when the Coulomb
repulsion $U$ is introduced. As expected, we have shown that the BI phase is transferred 
into the MI phase with increasing $U$. 
We have found that there are the following four cases for this transition to occur. 
i) The electronic phase transition occurs from a BI to a MI with trivial band topology, 
accompanied with closing the single-particle excitation gap (at non-TRIM points) at the same critical $U$ value. 
ii) The topological phase transition occurs from a TBI to a BI with the single-particle excitation gap 
closing at TRIM points, followed by the electronic phase transition from a BI to a topologically trivial MI 
with the single-particle excitation gap closing at non-TRIM points.
iii) The topological phase transition occurs within a MI from trivial to
nontrivial band topology
at the critical $U$ where the single-particle Green's function exhibits 
zeros, not poles, at the Fermi energy and at TRIM points. 
iv) In a wide range of $\theta$ values, 
the electronic phase transition occurs from a TBI to a MI 
without changing the band topology, 
where the single-particle excitation gap at the $K$ and $K'$ points is closed at the same critical 
$U$ value. 
Therefore, our calculations confirm the possibility of 
the MI phase with nontrivial band topology
in a $t_{2g}^5$ system with a honeycomb lattice structure.

\acknowledgements
The authors acknowledge S. Miyakoshi for fruitful discussion. 
The numerical calculations have been performed with 
the RIKEN supercomputer system (HOKUSAI GreatWave).
This work has been supported 
by Grant-in-Aid for Scientific Research from MEXT Japan 
under Grant No. 25287096 and also 
by RIKEN iTHES Project and Molecular Systems. 
K. S. acknowledges support from the JSPS Overseas Research Fellowships. 
T. S. acknowledges Simons Foundation for financial support (award no. 534160).

\appendix

\renewcommand{\thefigure}{A\arabic{figure}}
\renewcommand{\thetable}{A\arabic{table}}
\setcounter{table}{0}
\setcounter{figure}{0}

\section{Cluster perturbation theory} \label{app:cpt}

With the help of the exact diagonalization method based on the Lanczos algorithm~\cite{Morgan93},
we calculate the ground state $|\Psi_G \rangle$ with its energy $E_G$ for the $t_{2g}^5$ electron 
configuration in the six-site cluster under the open boundary conditions. 
Let $E_n^h$ ($E_n^e$) and $|\Psi_n^h \rangle$ ($|\Psi_n^e \rangle$) 
be the $n$-th eigenvalue and eigenstate of the cluster with 
the total number of electrons being one less (more) than that of the ground state. 
The cluster single-particle Green's function is given as 
\begin{equation}
G_{l \eta,l'\eta'}'(\zeta;\mu_c) = 
\sum_{m} \frac{ Q_{l \eta m}^e(Q_{l \eta' m}^e)^* } {\zeta-\epsilon_m^e-\mu_c} +
\sum_{n} \frac{ Q_{l \eta n}^h(Q_{l' \eta' n}^h)^* } {\zeta-\epsilon_n^h-\mu_c},
\label{cGf}
\end{equation} 
where $\zeta$ is complex frequency, $\epsilon_m^e = E_m^e-E_G$, $\epsilon_{n}^h = E_G - E_{n}^h$,
$Q_{l \eta m}^e = \langle \Psi_G |c_{l \eta} |\Psi_m^{e}\rangle$, and
$Q_{l \eta n}^h = \langle \Psi_n^{h}|c_{l \eta} |\Psi_G\rangle$~\cite{Zacher2002,Aichhorn2006}. 
$c_{l\eta}$ is the annihilation operator  
at site $l$ in the cluster and $\eta$ denotes both spin and orbital degrees of freedom. 
The chemical potential $\mu_c$ for the cluster is 
given as $\mu_c=\left( E_{0}^e-E_{0}^h \right) / 2$, where 
$E_0^h$ and $E_{0}^e$ are the minimum energy among $E_n^h$ and $E_n^e$, respectively.
Therefore, the Fermi energy is located in the middle of
the lowest one-electron and highest one-hole 
additional energy bands. 
Using the band Lanczos method~\cite{Freund}, we calculate 
$\epsilon_m^e$, $\epsilon_{n}^h$, $Q_{l \eta m}^e$, and $Q_{l \eta n}^h$ 
to obtain $G_{l \eta,l'\eta'}'(\zeta; \mu_c)$.

In the CPT~\cite{Senechal2002}, the lattice single-particle 
Green's function $\mathbf{G}(\zeta,\mathbf{K})$ of the supercell composed of the clusters 
is calculated as  
\begin{equation}
\mathbf{G}^{-1}(\zeta,\mathbf{K}) = 
\mathbf{G}'^{-1}(\zeta;\mu_c)-\mathbf{V}(\mathbf{K}),
\label{OCPT}
\end{equation}
where $\mathbf{G}'(\zeta;\mu_c)$ is the cluster Green's function given in Eq.~(\ref{cGf})
and $\mathbf{V}(\mathbf{K})$ is 
the Fourier transformation of the inter-cluster hopping matrix. 
Here, $\mathbf{K}$ is the momentum in the Brillouin zone of the supercell. 
Note that the lattice single-particle Green's function evaluated from the CPT 
sometimes fails to describe the total number of electrons correctly 
even when the cluster Green's function gives the correct number. 
This always happens when the electron-hole symmetry of $\mathbf{V}(\mathbf{K})$ is broken. 
To overcome this difficulty, here we adopt the VCA~\cite{Potthoff2003} 
with the chemical potential $\mu$ of the cluster
treated as a variational parameter. 
In this treatment, the cluster Green's function is calculated 
in Eq.~(\ref{cGf}) with $\mu_c$ replaced with $\mu$.
The additional term $(\mu-\mu_c)\mathbf{I}$ is also added in 
$\mathbf{V}(\mathbf{K})$ in order that
the replacement of the chemical potential does not change the overall Hamiltonian.
Thus, Eq.~(\ref{OCPT}) is modified as
\begin{equation}
\mathbf{G}^{-1}(\zeta,\mathbf{K}) = 
\mathbf{G}'^{-1}(\zeta;\mu)-\mathbf{V}_\mu(\mathbf{K}),
\label{CPT}
\end{equation}
where $\mathbf{V}_\mu(\mathbf{K})=\mathbf{V}(\mathbf{K})+(\mu-\mu_c)\mathbf{I}$.

The chemical potential $\mu$ is determined so as to satisfy the stationary condition 
of the grand potential function $\Omega(\mu)$, i.e., 
$\partial \Omega / \partial \mu |_{\mu=\mu^*}=0$, 
under the condition that the average number of electrons per site is 5. 
Here, the grand potential function $\Omega(\mu)$ at the zero temperature is  
given as
\begin{multline}
\Omega (\mu) = \Omega'(\mu)
+\frac{1}{2} \int_{BZ} d^2{\mathbf{K}} \tr\mathbf{V}_\mu(\mathbf{K})  \\
 - \int_0^{\infty} \frac{dx}{x} 
\int_{BZ} d^2{\mathbf{K}} \ln \vert 
\det \left[\mathbf{I}-\mathbf{V}_\mu(\mathbf{K})\mathbf{G}'(\textrm{i}x,\mu)\right]
\vert,
\end{multline}
where $\Omega'(\mu) = E_G-\mu N_t$ is the grand potential function of the cluster, 
$N_t$ is the total number of electrons in the cluster (i.e., $N_t=30$ for the six-site cluster), 
and $\int_{BZ} d^2{\mathbf{K}}\cdots$ refers to the integration of the momentum 
$\mathbf{K}$ over the Brillouin zone of the supercell.

\begin{figure}[t]
\centering
\includegraphics[width=0.95\columnwidth]{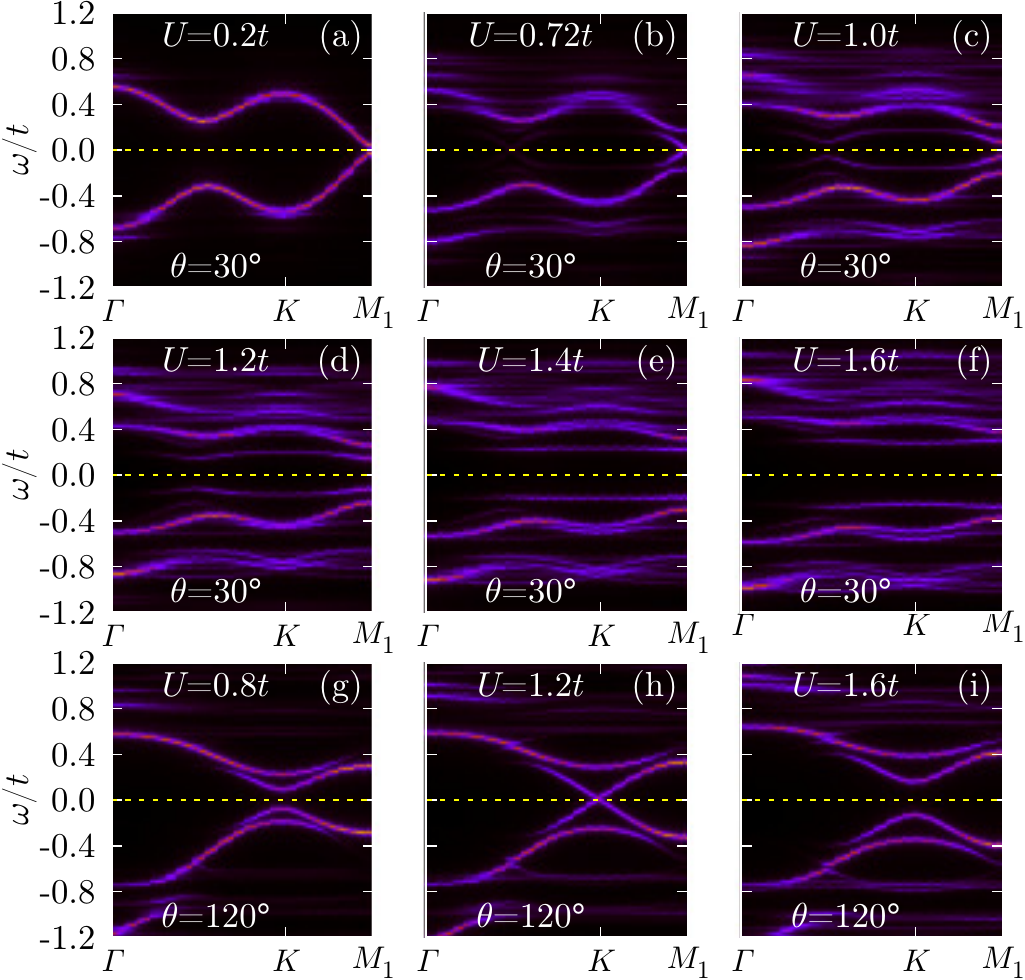}
\caption{
 Spectral functions $A\left(\omega,\mathbf{k}\right)$ for the same $U$ and $\theta$ values 
 used to obtain the energy 
 dispersions of the topological Hamiltonian $\mathbf{H}_{\rm T}(\mathbf{k})$ in Fig.~\ref{Fig_TH}.
 The corresponding $U$ and $\theta$ values are indicated in each figure. 
 We set that $\lambda=1.6t$ and $t_2=t_2'=t_3=\Delta_{\rm tr}=J_{\rm H}=0$. 
 The Fermi energy is located at $\omega=0$. 
}
\label{Fig_SFA}
\end{figure}

Note that the lattice single-particle Green's function $\mathbf{G}(\zeta,\mathbf{K})$ is given 
in terms of the supercell momentum $\mathbf{K}$. 
In general, the symmetry of the supercell can be different from 
that of the original honeycomb lattice. 
Therefore, in order to obtain the single-particle Green's function $\mathbf{G}(\zeta,\mathbf{k})$ 
at the momentum $\mathbf{k}$ in terms of the original honeycomb lattice, 
we periodize the Green's function as 
\begin{equation}
G_{\eta_j \eta'_{j'}} (\zeta,\mathbf{k})= \frac{1}{3} 
\sum_{l ,l'}  \delta_{j,l_2} \delta_{j',l'_2} 
G_{l \eta,l' \eta'} (\zeta,\mathbf{K}) 
e^{\textrm{i}\mathbf{k} \cdot  ( \mathbf{r}_l-\mathbf{r}_{l'})},
\end{equation}
where $\eta_j$ in the left hand side is referred to as state $\eta$ at the $j$-th ($j=0,1$) base in the 
unit cell of the original honeycomb lattice, and 
$l_2$ is the remainder after dividing $l$ by 2 (i.e., $l_2 = l\mod2$). 
$\mathbf{r}_l$ is the lattice vector of the unit cell of the honeycomb lattice within the 
cluster that contains site $l\,(=0,1,\dots,5)$. 
The supercell momentum $\mathbf{K}$ that corresponds to the momentum $\mathbf{k}$ 
can be obtained by properly subtracting from $\mathbf{k}$ a reciprocal lattice vector $\mathbf{k}_s$ 
of the supercell. 
Finally, the periodized spectral function $A\left(\mathbf{k},\omega\right)$ can 
be evaluated as
\begin{equation}
A\left(\mathbf{k},\omega\right) = 
-\frac{1}{\pi}\sum_{j,\eta} \textrm{Im} 
 G_{\eta_j,\eta_j}(\omega+\textrm{i}\delta,\mathbf{k}),
\label{Eq_SF}
\end{equation}
where $\omega$ is real frequency 
and $\delta$ is the broadening parameter taken as $\delta=0.008t$ in our calculations.


\section{Spectral functions}\label{app:sf}

\begin{figure}[t]
\centering
\includegraphics[width=0.95\columnwidth]{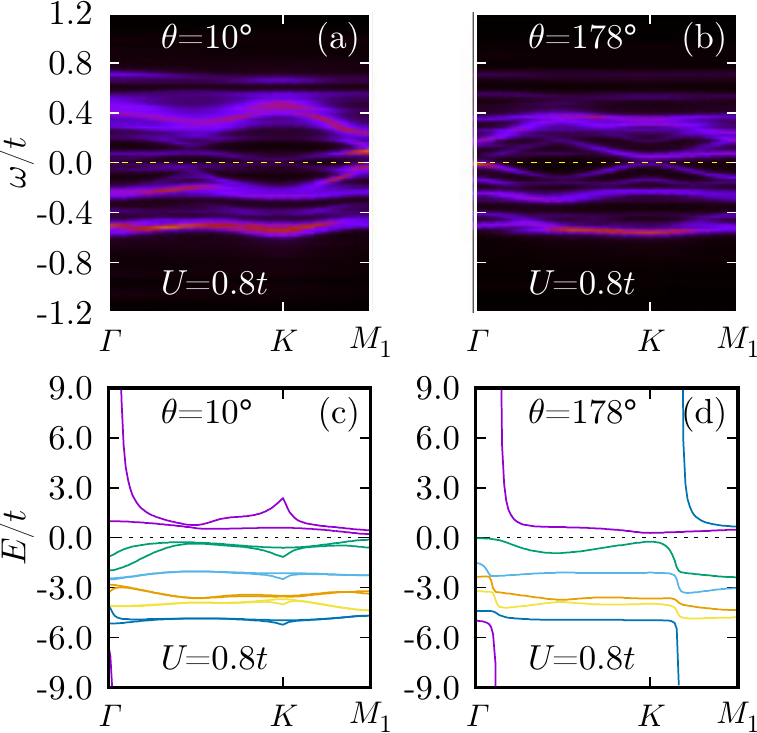}
\caption{
 (a, b) Spectral functions $A\left(\mathbf{k},\omega\right)$ of the interacting system and 
 (c, d) energy dispersions of the topological Hamiltonian $\mathbf{H}_{\rm T}(\mathbf{k})$ 
 for $\theta=10^\circ$ and $178^\circ$ indicated in the figures. 
 We set that $U=0.8t$, $\lambda=1.6t$, and $t_2=t_2'=t_3=\Delta_{\rm tr}=J_{\rm H}=0$. 
 These parameters are in the hatched areas of the topological phase diagram shown in Fig.~\ref{Fig_U}(a). 
}
\label{Fig_BD}
\end{figure}

Figure~\ref{Fig_SFA} shows the spectral functions calculated from Eq.~(\ref{Eq_SF}) 
for the same parameters used to obtain the energy dispersions of 
the topological Hamiltonian $\mathbf{H}_{\rm T}(\mathbf{k})$ in Fig.~\ref{Fig_TH}. 
When $\theta=30^\circ$, we can observe that, as $U$ increases, subbands 
with weak intensity appear inside the main bands with dominant intensity.
On the other hand, when $|t_1'|>|t_1|$ as in the case of $\theta=120^\circ$, 
the overall shapes of spectral functions near the Fermi energy resemble 
those for the Kane-Mele-Hubbard model 
with a finite SOC (see Refs.~\onlinecite{Yu2011,Grandi2015}). 
In this case, as shown in Figs.~\ref{Fig_SFA}(g)--\ref{Fig_SFA}(i),
the conduction and valence bands around the $K$ and $K'$ points 
are split into two subbands, each of which exhibits similar spectral weight.

\section{Breakdown of the topological Hamiltonian}\label{app:th}

We find that the topological Hamiltonian $\mathbf{H}_{\rm T}(\mathbf{k})$ obtained in our calculations 
is sometimes broken down for the particular parameter regions specially 
when $t_1$ is predominant. 
For example, the topological Hamiltonian for $\theta=10^\circ$ and $U=0.8t$ 
shown in Fig.~\ref{Fig_BD}(c) does not preserve the IS and TRS as it should. 
The topological Hamiltonian for $\theta=178^\circ$ and $U=0.8t$
shown in Fig.~\ref{Fig_BD}(d) exhibits several singularities, although 
it preserves the correct symmetry. 
In these cases, the $Z_2$ topological invariant based on the
topological Hamiltonian is not well defined. 
One possibility of these kinds of breakdown is due to the failure of precise 
numerical calculations. 
We have found that the convergence of the ground state for these parameters 
as in Fig.~\ref{Fig_BD}(c)
is much poorer than that for other parameters. 
On the other hand, the convergence of the ground state for the parameters such as 
the case in Fig.~\ref{Fig_BD}(d) is almost similar to that for other parameters where the topological Hamiltonian 
is well defined. 
Thus, in this case, the numerical error for the calculation of the ground state does not seem serious. 
To resolve these difficulties, more precise numerical analysis is required.


\end{document}